\documentclass[reprint,amsmath,amssymb,amsfonts,twocolumn,nofootinbib]{revtex4-1}

\usepackage{bm,graphicx,mathrsfs}
\usepackage{graphicx}
\usepackage{epsfig}
\usepackage{amsmath,bbm}
\usepackage{amsfonts,amssymb}
\usepackage{times}
\usepackage{dsfont}
\usepackage{enumitem}  

\newcommand{\ket}[1]{|#1\rangle}
\newcommand{\bra}[1]{\langle#1|}

\begin{document}

\title{Autonomous Quantum Error Correction and Application to Quantum Sensing with Trapped Ions}

\author{F. Reiter$^{1,2,3}$, A. S. S\o{}rensen$^{4}$, P. Zoller$^{1,2}$, and C. A. Muschik$^{1,2}$}
\affiliation{$^{1}$Institute for Theoretical Physics, University of Innsbruck, A-6020 Innsbruck, Austria}
\affiliation{$^{2}$Institute for Quantum Optics and Quantum Information of the Austrian Academy of Sciences, A-6020 Innsbruck, Austria}
\affiliation{$^{3}$Harvard University, Department of Physics, Cambridge, MA 02138, USA}
\affiliation{$^{4}$Niels Bohr Institute, University of Copenhagen, Blegdamsvej 17, DK-2100 Copenhagen, Denmark}
\date{\today}

\begin{abstract}
Quantum-enhanced measurements hold the promise to improve high-precision sensing ranging from the definition of time standards to the determination of fundamental constants of nature. However, quantum sensors lose their sensitivity in the presence of noise. To protect them, the use of quantum error-correcting codes has been proposed. Trapped ions are an excellent technological platform for both quantum sensing and quantum error correction.
Here we present a quantum error correction scheme that harnesses dissipation to stabilize a trapped-ion qubit. In our approach, always-on couplings to an engineered environment protect the qubit against spin or phase flips. Our dissipative error correction scheme operates in a fully autonomous manner without the need to perform measurements or feedback operations. We show that the resulting enhanced coherence time translates into a significantly enhanced precision for quantum measurements. Our work constitutes a stepping stone towards the paradigm of self-correcting quantum information processing.
\end{abstract}

\maketitle

Quantum noise processes impose strong limitations on devices that take advantage of quantum mechanics, such as quantum computers \cite{Ladd}, quantum networks~\cite{Kimble}, quantum simulators~\cite{Lloyd}, and quantum-enhanced sensors \cite{Ramsey, Huelga, Giovanetti}.
The quest to find viable strategies for mitigating errors is thus an essential prerequisite for the development of quantum technologies and has led to techniques for quantum error correction~\cite{Shor, Nielsen, Fernando}.
The application of quantum error-correcting codes to improve quantum sensing protocols \cite{Dur, Arrad, Kessler, Ozeri, Sekatski, Unden} represents a young research direction. This approach is complementary to control methods \cite{Control1,Control2}; however, it addresses the same challenge of improving the sensitivity of quantum measurements in the presence of noise \cite{Acin, Toth}.
In particular, trapped-ion systems \cite{LBMW} have proven to be an excellent platform for high-precision measurements~\cite{Precision-Ions-1,Precision-Ions-2,Precision-Ions-3}, as well as for the realization of quantum error-correcting codes \cite{ExperimentWineland, ExperimentBlatt}. Still, the combination of the two techniques has not yet been demonstrated.

Harnessing dissipative processes by engineering the coupling of a system to an environment or reservoir ~\cite{PCZ, PHBK, Kraus, Diehl, VWC, Metelmann, Morigi, RRS} provides a route for processing quantum information alternative to relying on unitary gate operations \cite{CiracZoller, ExperimentWineland, ExperimentBlatt, Monz}. Such reservoir engineering techniques have for example been successfully applied for state preparation \cite{Krauter, Barreiro, Lin, Shankar, Kienzler}. Dissipative schemes for preparing entangled resource states have been shown to have
advantages over standard methods, leading for example to a better use of resources~\cite{KRS}
and extending entanglement-lifetimes by stabilizing the target state~\cite{Krauter}.

\begin{figure}[t]
\centering
\includegraphics[width=8.6cm]{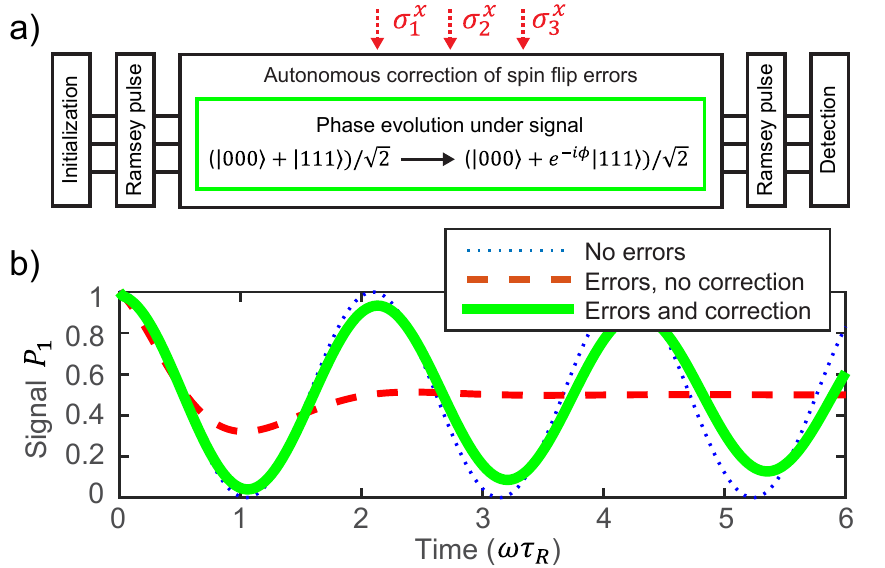}
\caption{
Application of the proposed error correction scheme to quantum metrology. (a) Ramsey scheme. From an initial state $\ket{000}$, a Ramsey pulse prepares $(\ket{000} \!+ \!\ket{111})/\sqrt{2}$. Under $H \!=\! (\omega/2) \sum_{j=1}^N \sigma^z_j$, the state evolves to $(\ket{000} + e^{-i\phi}\ket{111})/\sqrt{2}$ during a Ramsey time $\tau_{\rm R}$. Rotation by a second Ramsey pulse and measurement of the population $P_1$ of one ion in the $\{\ket{0},\ket{1}\}$ basis allows deducing $\omega$ according to $P_1=\cos^2(\phi/2)$ with $\phi=3\omega\tau_{\rm R}$.
(b) In the absence of errors, the Ramsey measurement has full fringe contrast (dotted blue line). Spin flips at a rate $\Gamma = \omega/2$ damp out the oscillations (dashed red line). Applying the presented scheme for error correction protects the evolution and restores the fringe contrast (solid green line, assuming a sideband coupling of $G = 5000 ~ \Gamma$). A detailed discussion of the achievable sensitivity is given in the text.
}
\label{FigMetrology}
\end{figure}

Employing dissipation for quantum error correction takes this idea further since it requires the stabilization of an \textit{unknown} state (i.e., of a manifold of states). The idea of dissipative error correction has attracted considerable interest \cite{Paz1998, Ahn2002, Sarovar2005, Oreshkov2007, Mabuchi2009, Pastawski, Ippoliti2014, Fujii2014}. The challenge of implementing this strategy by engineering suitable dissipative processes in concrete experimental systems has recently led to theoretical proposals for superconducting circuits~\cite{Leghtas, Kapit1, Kapit2, Cohen, Freeman, ChannelConstruction}, as well as first experimental efforts towards the realization of building blocks required for dissipative quantum error correction~\cite{Leghtas2015}. Despite their central roles in both quantum information processing and quantum metrology, such achievements have not been made with trapped ions.

In this work, we address this challenge by combining the paradigms of dissipative quantum error correction and trapped-ion quantum information processing to a novel scheme for quantum error-corrected metrology (see Fig.~\ref{FigMetrology}).
Standard error-correcting schemes \cite{Nielsen} entail classical apparatuses to perform measurements and feedback operations on the quantum system. In contrast, our scheme neither relies on time-dependent unitary quantum gates, nor requires macroscopic measurements or feedback operations. Instead, always-on couplings to an engineered environment autonomously correct for spin or phase flips at a microscopic level. The resulting protection of a qubit against noise results in a significant enhancement of its lifetime, and hence in a substantial improvement of quantum measurements, as can be seen from Fig. \ref{FigMetrology}.

\begin{figure}[t!]
\centering
\includegraphics[width=8.6cm]{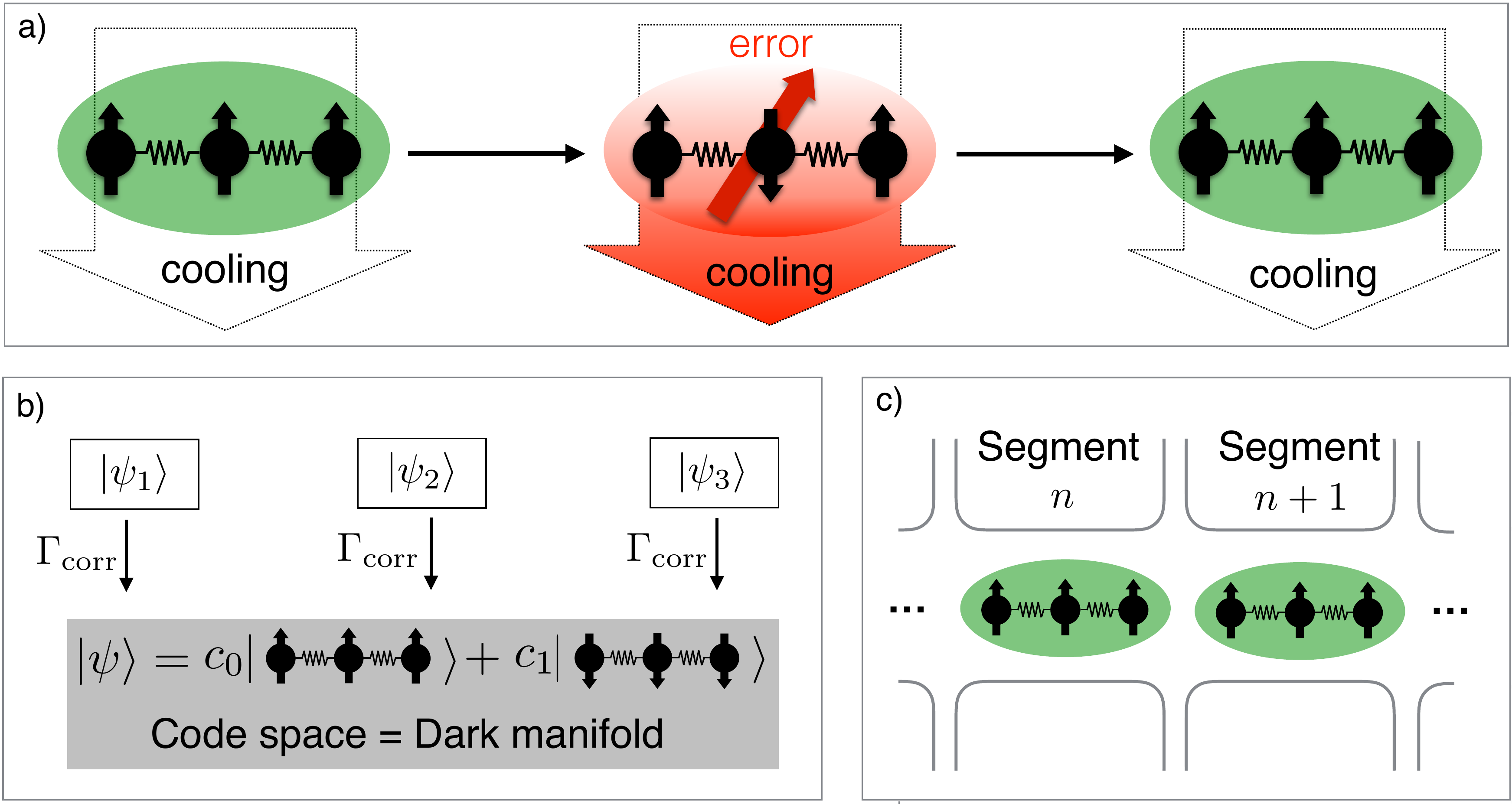}
\caption{
Dissipative realization of a three-qubit code in trapped ions. The unknown state $\ket{\psi} = c_0 \ket{000}+ c_1 \ket{111}$ is encoded in the internal states of three ions and protected against single spin flips by means of engineered couplings to a quantum reservoir that consists of motional modes. (a) Bit flip errors are continuously removed from the system through cooling of the motional modes. (b) Compromised states $\ket{\psi_j}=\sigma_j^x\ket{\psi}, j=1,2,3$ are driven back to the logical subspace at a rate $\Gamma_{\rm corr}$, while the code space is a dark manifold of the engineered dissipative processes. (c) The scheme can be scaled up to several logical qubits using a segmented trap (see App.~\ref{App_Metrology_Scalability}).
}
\label{FigConcepts}
\end{figure}

The proposed scheme allows for the realization of a repetition code~\cite{Shor}, where a logical qubit is encoded in a three-particle entangled state. Dissipative processes are designed such that the code space is a steady state dark manifold, as shown in Fig. \ref{FigConcepts}. Errors take the quantum state out of this subspace, which causes engineered dissipative processes to become resonant and thereby coherently correct the error by a generalized optical pumping process. Once the error is corrected, the engineered dissipative processes are shifted out of resonance such that they cease to act on the system. As we show below, this allows one to stabilize a qubit continuously against single-qubit spin flips or phase flips or against correlated noise.

Adding to the potential of this approach for quantum information processing we demonstrate that the proposed scheme can be applied for improving the sensitivity of quantum sensing protocols. Here our scheme provides a blueprint for quantum error correction enhanced metrology in trapped-ion systems based on current experimental means. We analyze the applicability of our scheme in the context of a paradigmatic measurement setting with trapped ions and show that the attainable sensitivity can be significantly enhanced for realistic experimental parameters.
Due to its autonomous character, our scheme provides a stepping stone towards a new paradigm of self-correcting quantum systems that can be realized with current technology.

In summary, our scheme allows one to realize an autonomously protected qubit by coupling the internal degrees of freedom of a system of trapped ions to an environment consisting of cooled motional modes (see Fig.~\ref{FigConcepts}). Our analysis shows that this approach can be used for quantum metrology and will enable experiments with ions that take dissipative quantum information processing to a new level.

The remainder of the paper is organized as follows. We start with a description of the dissipative error correction protocol and proceed with presenting the setup and the key mechanisms of our scheme. To assess the performance of this protocol, we analyze a simplified effective model, compare it with a numerical simulation and discuss the effect of experimental imperfections. Finally, we demonstrate the application of our scheme in the context of quantum metrology, where we use the error correction protocol to protect Ramsey measurements against undesired noise processes.

\section{Error-correcting protocol}

We consider a three-qubit repetition code, where a logical qubit is encoded in three physical qubits,
\begin{align*}
\ket{\psi} = c_0 \ket{0}_L + c_1 \ket{1}_L = c_0 \ket{000}+ c_1 \ket{111},
\end{align*}
with logical qubit states $\ket{0}_L = \ket{000}$ and $\ket{1}_L = \ket{111}$. This encoding allows one to protect the quantum state against single bit flips on the physical qubits, which lead to the single-error states
\begin{align}
\ket{\psi_j} &= \sigma^x_j \ket{\psi}, ~ ~ ~ j=1,2,3,
\label{EqPsij}
\end{align}
where the Pauli operator $\sigma^x_j = \ket{0}_j \bra{1} + \ket{1}_j \bra{0}$ acts on the $j^{\rm th}$ qubit. As explained below, a majority vote allows for the restoration of the state $\ket{\psi}$.
Later we will also consider the correction of single-qubit phase flips and correlated noise. In principle it is also possible to convert this scheme into a protocol correcting phase errors and then to extend it to a nine-qubit code~\cite{Shor}, which is capable of correcting any type of single-qubit error. However, a full error-correcting code is not compatible with our application for quantum sensing, since it would remove both errors and the signal to be measured from the dynamics. In the following, we describe the noisy dynamics~\cite{Gardiner} of a quantum system by a master equation $\dot{\rho}=\mathcal{L}(\rho)$ with a Liouvilllian $\mathcal{L}$ of Lindblad form,
\begin{align}
\mathcal{L_{\rm noise}}(\rho) &= \sum_k \mathcal{D}[L_k](\rho),
\label{EqLiouvillianNoise}
\\ \nonumber
\mathcal{D}[L_k](\rho) &= L_k \rho L_k^\dagger - \frac{1}{2} \left( L_k^\dagger L_k \rho + \rho L_k^\dagger L_k \right),
\end{align}
with dissipators $\mathcal{D}[L_k]$ and jump operators $L_k$. Local spin flips are described by jump operators
\begin{align}\label{EqSpinFlip}
L_{x_j} = \sqrt{\Gamma} \sigma^x_j, ~ ~ ~ j=1,2,3,
\end{align}
where $\Gamma$ is the rate at which a bit flip occurs.
Single bit flip errors are corrected by implementing a majority vote: To correct errors on, for example, the second qubit, we interrogate the two-body stabilizer operators $S_{12} = \sigma^z_{1} \sigma^z_{2}$ and $S_{23} = \sigma^z_{2} \sigma^z_{3}$, which involve the second qubit and its neighbors, qubit $1$ and qubit $3$. The code state $\ket{\psi}$ is an eigenstate of these operators with eigenvalue $+1$, while the single error state $\ket{\psi_2}$ is an eigenstate with eigenvalue $-1$.
If a spin flip has occurred, the resulting state $\ket{\psi_2}$ of the system thus violates both stabilizers, $S_{12} \ket{\psi_2} = S_{23} \ket{\psi_2} = - \ket{\psi_2}$. Conditioned on this result, a spin flip $\sigma^x_2$ is applied to qubit $2$, and the original state $\ket{\psi}$ is restored. Errors on the first and third qubits are corrected in an analogous fashion.
This recovery protocol can be implemented in a continuous and autonomous manner by realizing the dynamics
\begin{align*}
\mathcal{L}_{\rm corr} &= \mathcal{D}[L^{x_1}_{\rm corr}] + \mathcal{D}[L^{x_2}_{\rm corr}] + \mathcal{D}[L^{x_3}_{\rm corr}],
\end{align*} 
with jump operators  of the form
\begin{align}\label{EqLcorrX2}
L^{x_2}_{\rm corr} &= \sqrt{\Gamma_{\rm corr}} \sigma^x_{2} \frac{\mathbbm{1} - \sigma^z_{1} \sigma^z_{2}}{2} \frac{\mathbbm{1} - \sigma^z_{2} \sigma^z_{3}}{2}.
\end{align}
Here, we have $\Gamma_{\rm corr}$ as the correction rate, $\sigma^x_2$ as the correcting spin flip and two ``interrogation parts'' of the form $(\mathbbm{1}-S)/2$. Each of them interrogates a stabilizer $S$, yielding $0$ if the qubits are of the same value and $\mathbbm{1}$ if they are different. This realizes the described majority vote: If both stabilizers $S_{12}$ and $S_{23}$ are violated, an action $L_{\rm corr}^{x_2} \sim \sigma^x_2$ is realized. Correcting operators for qubits 1 and 3 can be written analogously:
\begin{align}
L^{x_1}_{\rm corr} &= \sqrt{\Gamma_{\rm corr}} \sigma^x_{1} \frac{\mathbbm{1} - \sigma^z_{1} \sigma^z_{2}}{2} \frac{\mathbbm{1} - \sigma^z_{1} \sigma^z_{3}}{2},
\label{EqLcorrX1}
\\
L^{x_3}_{\rm corr} &= \sqrt{\Gamma_{\rm corr}} \sigma^x_{3} \frac{\mathbbm{1} - \sigma^z_{1} \sigma^z_{3}}{2} \frac{\mathbbm{1} - \sigma^z_{2} \sigma^z_{3}}{2}.
\label{EqLcorrX3}
\end{align}
For the physical implementation, it will be useful to translate the conditional jump operators in Eqs. \eqref{EqLcorrX2}--\eqref{EqLcorrX3} into the form
\begin{align}\label{EqLcorr}
L^{x_j}_{\rm corr} = \sqrt{\Gamma_{\rm corr}} ( \sigma^-_j P_{n_1=1} + \sigma^+_j P_{n_0=1}), ~ ~ ~ j=1,2,3,
\end{align}
where $\sigma^+_j$ ($\sigma^-_j$) are the raising (lowering) operators on qubit $j$. $P_{n_k=n}$ are projectors on the states with $n$ qubits in state $\ket{k}$, e.g., $P_{n_1=1} = \ket{100}\bra{100} + \ket{010}\bra{001} + \ket{001}\bra{001}$ and $P_{n_0=1} = \ket{011}\bra{011} + \ket{101}\bra{101} + \ket{110}\bra{110}$.
Conditional jump operators containing an interrogation part $(\mathbbm{1}-S)/2$ were proposed in~\cite{Kraus} and implemented in~\cite{Barreiro} using a sequence of quantum gates to dissipatively generate Bell pairs.
In the following, we show how the conditional jump operator performing a majority vote in Eq. \eqref{EqLcorr} can be implemented in a time-continuous fashion in a system of trapped ions.

\begin{figure}[t!]
\centering
\includegraphics[width=8.6cm]{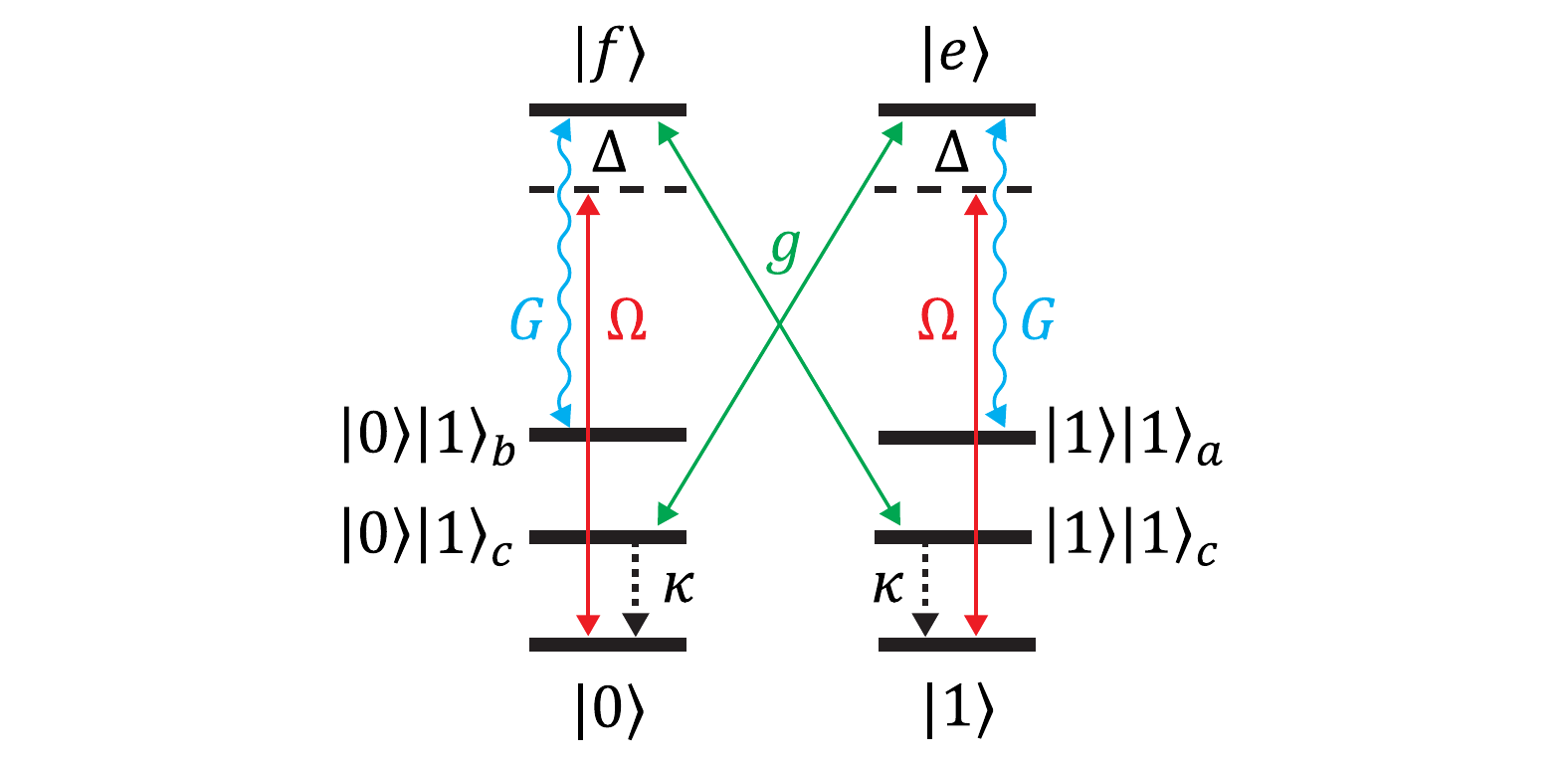}
\caption{
Setup. The setup consists of three system ions and two oscillator modes $a$ and $b$. For each physical error we require a coupling of the system ions to an ancillary system $c_j$, which can either be an oscillator mode or an ancilla ion. In the figure this additional degree of freedom is denoted by a level $\ket{1}_c$. The ions are assumed to have two ground levels, $\ket{0}$ and $\ket{1}$, and two excited levels, $\ket{e}$ and $\ket{f}$. The transitions are excited by a driving (weak carrier) field $\Omega$ and strongly coupled to two motional sidebands $a$ and $b$ with coupling constant $G$. Excitations of the ions ($\ket{e}$, $\ket{f}$) are coherently transferred to the ancillary systems by coherent couplings $g$ and subsequently removed by cooling/reset with a rate $\kappa$.
}
\label{FigSetup}
\end{figure}

\section{Setup}

We consider a setup consisting of three ions in a trap with identical levels and couplings as shown in Fig. \ref{FigSetup}.
The logical qubit states $|0\rangle$ and $|1\rangle$ are encoded in internal electronic levels of the ions.
In addition, we consider two excited levels, $\ket{e}$ and $\ket{f}$.
The internal levels are coupled to two motional modes of the ion chain, $a$ and $b$, by applying appropriate laser fields \cite{CiracZoller, LBMW}.
The four states are assumed to be at least meta-stable and the modes are assumed to be cooled to the ground state. In a suitable rotating frame, the free Hamiltonian of the system is given by
\begin{align*}
H_{\rm free} = \Delta \sum_{j=1}^3 \left( \ket{e}_j \bra{e} + \ket{f}_j \bra{f} \right) + \delta \left( a^\dagger a + b^\dagger b\right).
\end{align*}
As will be explained below, the selectivity of the correction processes (i.e., spin flips $\sigma_j^x$) to the single error states $\ket{\psi_j}$ will be achieved by a suitable choice of the detunings $\Delta$ and $\delta$.
To interrogate the system, we apply weak carrier drives on the transitions $\ket{1} \leftrightarrow \ket{e}$ and $\ket{0} \leftrightarrow \ket{f}$,
\begin{align}
H_{\rm drive} = \frac{\Omega}{2} \sum_{j=1}^N \left( \ket{e}_j \bra{1} + \ket{f}_j \bra{0} \right) + H.c.,
\label{EqHdrive}
\end{align}
and use sideband couplings that couple the transition $\ket{e} \rightarrow \ket{1}$ to mode $a$ and the transition $\ket{f} \rightarrow \ket{0}$ to mode $b$,
\begin{align*}
H_{\rm osc} = G \sum^{3}_{j=1}\left( a^\dagger \ket{1}_j \bra{e} + b^\dagger \ket{0} _j\bra{f} \right) + H.c.
\end{align*}
This description is valid within the Lamb-Dicke approximation \cite{LambDicke}. The coupling strength $G$ is assumed to be strong compared to the other rates in the system. The combination of the couplings $H_{\rm drive}$ and $H_{\rm osc}$, and the detunings in $H_{\rm free}$ will be used to identify ions which, after an error, reside in the state $\ket{1}$ ($\ket{0}$) and drive them to the state $\ket{e}$ ($\ket{f}$).
While the modes $a$ and $b$ are needed for the interrogation of the state of the system, we also require couplings to additional ancillary systems to remove errors from the system ions.

Eq. \eqref{EqLcorr} shows that the recovery process consists of two parts, one of which transfers states with a single qubit in state $\ket{1}$ back to $\ket{000}$, e.g., $\ket{001}\rightarrow\ket{000}$. The other process restores the state $\ket{111}$ from states with a single qubit in state $\ket{0}$, e.g. $\ket{110}\rightarrow\ket{111}$. For implementing the operators in Eq. \eqref{EqLcorr} it is important to maintain the coherence between these two parts while removing the errors.
Thus, individual uncorrelated dissipation such as decay of $\ket{e}$ and $\ket{f}$ by spontaneous emission does not suffice. Instead, we use an engineered cooling process which is a combination of a mapping of the errors from the system ions to ancillary systems and a subsequent dissipative process to remove the errors as explained below. These auxiliary systems can be either additional ions or additional motional modes. The exact level structure is not important, the only requirement is that the first excited level needs to be strongly damped, thus effectively resulting in a two-level system. In the following, we focus on implementations based on the use of additional motional modes subject to sympathetic cooling~\cite{Lin}. Implementations based on the use of ancilla ions can be described in an analogous fashion by replacing the bosonic operator $c_j$ in Eq.~\eqref{EqHanc} and Eq.~\eqref{EqDecay} by the Pauli operator $\sigma_j^{-}$. In this case, the required dissipative mechanism $L_{\sigma^-_j} = \sqrt{\kappa} \sigma^-_j$ acting on each ancilla system can be realized by continuously resetting the state of the ancilla ions by means of optical pump fields~\cite{Barreiro}.
 
For the simultaneous correction of each source of errors (spin flips acting independently on qubits 1, 2,  and 3), the excitations of the system ions are coherently mapped onto motional modes $c_j$. This is achieved by sideband couplings
\begin{align}
H_{\rm anc} = g \sum_{j=1}^3 e^{i \delta_c t} c_j^\dagger \left( \ket{0}_j \bra{e} + \ket{1}_j \bra{f} \right) + H.c.,
\label{EqHanc}
\end{align}
with the coupling strength $g$ and the detuning $\delta_c$.
The Hamiltonian of the overall system is given by
\begin{align*}
H_{\rm total} = H_{\rm free} + H_{\rm drive} + H_{\rm osc} + H_{\rm anc}.
\end{align*}

The resulting excitations in the motional modes are removed by a dissipative process,
\begin{align}\label{EqDecay}
L_{c_j} = \sqrt{\kappa} c_j,
\end{align}
for which we assume a large cooling rate $\kappa \gg g, \delta_c$ so that we can adiabatically eliminate $c_j$ from the dynamics \cite{EO}. This yields the engineered cooling operators
\begin{align}
L_{{\rm eng},j} &= \sqrt{\kappa_{\rm eng}} (\ket{0}_j \bra{e} + \ket{1}_j \bra{f}),
\label{EqEngDecay}
\end{align}
with an engineered cooling rate $\kappa_{\rm eng} = g^2 / \kappa$. The minor shift of the levels by $H_{\rm anc}$ can be neglected and the Hamiltonian becomes
\begin{align}\label{EqHtot}
H_{\rm total, eng} = H_{\rm free} + H_{\rm drive} + H_{\rm osc}.
\end{align}
As opposed to other sources of noise such as spontaneous emission and decay of the vibrational modes, the engineered decay in Eq. \eqref{EqEngDecay} preserves the coherence between $\ket{e}$ and $\ket{f}$ ($c_0\ket{e}+c_1\ket{f}\rightarrow c_0\ket{0}+c_1\ket{1}$).
\\

\begin{figure}[t]
\centering
\includegraphics[width=8.6cm]{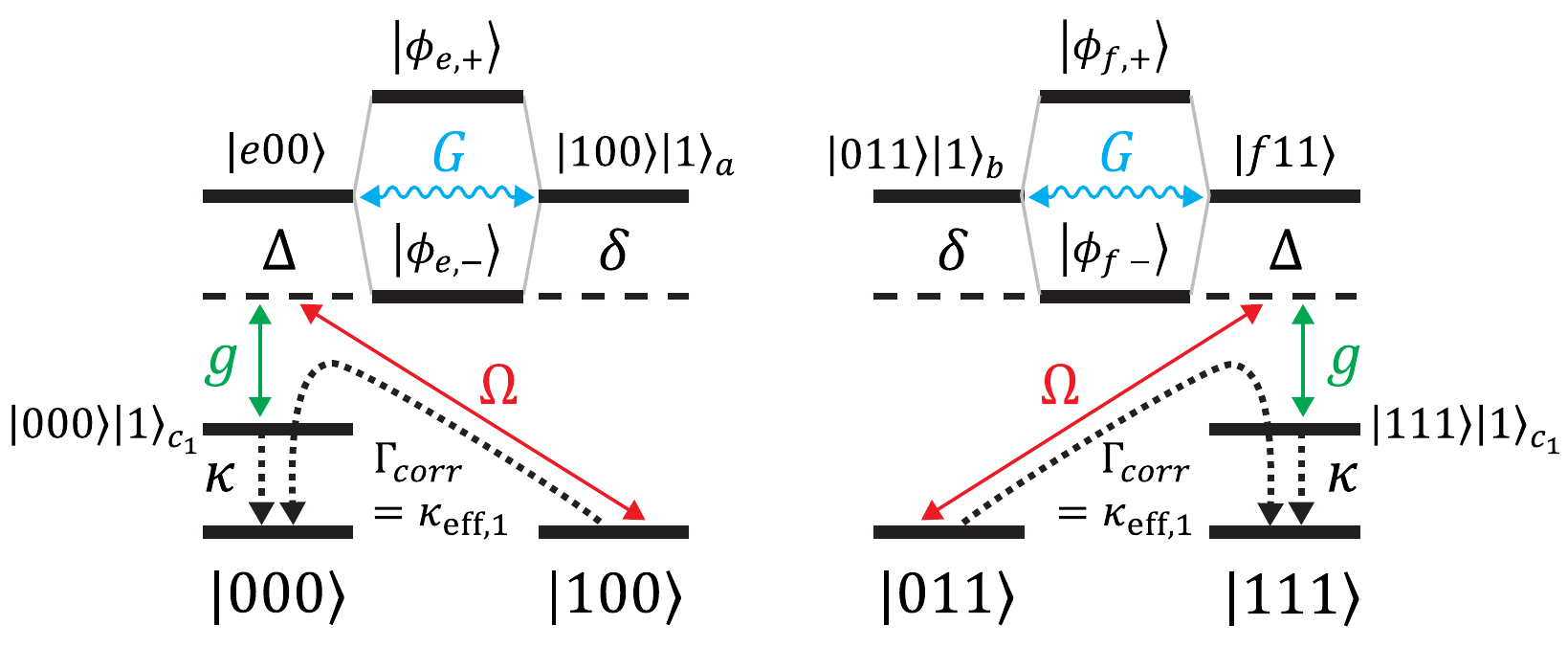}
\caption{
Error correction mechanism. The states after a spin flip on ion $1$, $\ket{100}$ and $\ket{011}$, are off-resonantly excited to the excited states $\ket{e00}$ (left) and $\ket{f11}$ (right). The strong sideband coupling $G$ results in the formation of the dressed states $\ket{\phi_{e/f,\pm}}$. For $\Delta=\delta=G$, the lower dressed states $\ket{\phi_{e/f,-}}$ are in resonance with the drives from the single-error states,
allowing for their selective excitation. The excitations in $\ket{e}$ and $\ket{f}$ are transferred from the ion to the ancillary system $c_1$ by a coherent coupling, $g$, and removed by cooling $\kappa$, maintaining the coherence between the two paths $\ket{100}\rightarrow\ket{000}$ and $\ket{011}\rightarrow\ket{111}$. Correction of spin flips on other ions $j$ is performed in the same way, utilizing the ancillary systems $c_j$.
}
\label{FigMechanism}
\end{figure}

\begin{figure*}[t]
\centering
\includegraphics[width=17.2cm]{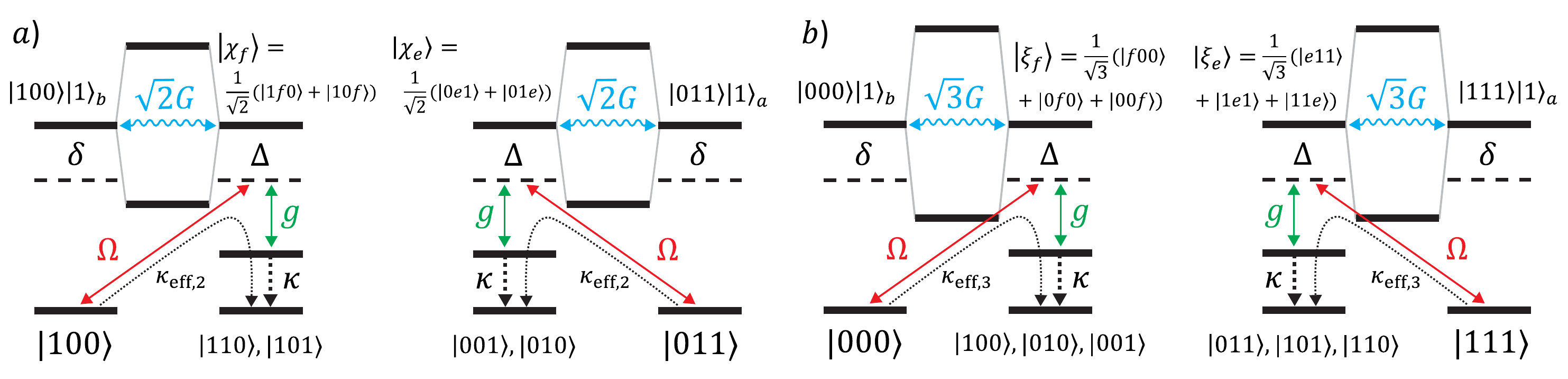}
\caption{
Intrinsic error processes.
(a) Excitation of erroneous but correctable states, e.g., $\ket{100}$ and $\ket{011}$ can lead to excited states such as $\ket{\xi_f}=(\ket{1f0}+\ket{10f})/\sqrt{2}$ instead of $\ket{e00}$ and $\ket{\xi_e}=(\ket{0e1}+\ket{01e})/\sqrt{2}$ instead of $\ket{f11}$. Since their dressed states with the oscillator-excited states $\ket{100}\ket{1}_b$ and $\ket{011}\ket{1}_a$ reside at detunings $|\Delta| = \sqrt{2} G$, such processes are off-resonant with respect to the drive, resulting only in weak excitation. These processes, however, lead to a leakage of population from the subspace of correctable states since they generate double errors, e.g. $\ket{100} \rightarrow (\ket{110} + \ket{101})/\sqrt{2}$ and $\ket{011} \rightarrow (\ket{010} + \ket{001})/\sqrt{2}$, which cannot be corrected by our protocol.
(b) Excitation of the logical states $\ket{000}$ and $\ket{111}$ leads to excited states $\ket{\chi_f}$ and $\ket{\chi_e}$ that couple to oscillator-excited states with strength $\sqrt{3} G$ and thus form dressed states at energies $\pm \sqrt{3} G$. These states are driven off-resonantly by the weak drive, which has a detuning $|\Delta| = G$. Such off-resonant excitations and subsequent engineered cooling processes transfer code states to the manifold with one error and can thus be recovered by the correction mechanism in the same manner as the physical error.
}
\label{FigErrors}
\end{figure*}

\section{Working mechanism}

We use the above couplings to implement the dissipative error-correcting dynamics in Eq. \eqref{EqLcorr}. Fig.~\ref{FigMechanism} illustrates the correction of the encoded qubit after a spin flip on qubit $1$. Spin flips on qubits $2$ and $3$ are corrected analogously. Starting from the single-error state after a bit flip on qubit $1$ (see Eq. \eqref{EqPsij}),
\begin{align}
\ket{\psi_1} = c_0 \ket{100} + c_1 \ket{011},
\label{EqPsi1}
\end{align}
the weak drive $H_{\rm drive}$ couples $\ket{100}$ to $\ket{e00}$ and $\ket{011}$ to $\ket{f11}$.
This excitation is a priori off-resonant due to the detunings in $H_{\rm free}$. We now use the oscillator coupling $H_{\rm osc}$ to create resonances in the excited state manifold which are selective on the number of qubits in $\ket{0}$ or $\ket{1}$ in the initial state: $H_{\rm osc}$ couples $\ket{e00}$ back to $\ket{100}$ by exciting oscillator $a$, resulting in the state $\ket{100}\ket{1}_a$. This coupling between $\ket{e00}$ and $\ket{100}\ket{1}_a$ leads to the formation of dressed states,
\begin{align*}
\ket{\phi_{e,\pm}} = c_{e,\pm} \ket{e00} \pm c_{a,\pm} \ket{100}\ket{1}_{a},
\end{align*}
which are indicated in Fig. \ref{FigMechanism}.
For $\Delta = \delta = G$, the lower dressed state resides at the detuning $\Delta_{1,-} = 0$ (the upper one at $\Delta_{1,+} = 2 G$), such that it is in resonance with the drive $H_{\rm drive}$ from $\ket{100}$. Hence, $\ket{100}$ is excited to $\ket{\phi_{e,-}}$.

Simultaneously, the $\ket{011}$-part in Eq. \eqref{EqPsi1} is excited to the state $\ket{f11}$. The excitation $\ket{f}$ is coupled to oscillator $b$ by $H_{\rm osc}$, which results in the formation of the dressed states
\begin{align*}
\ket{\phi_{f,\pm}} = c_{f,\pm} \ket{f11} \pm c_{b,\pm} \ket{011}\ket{1}_{b},
\end{align*}
with the energies $\Delta_{1,\pm}$.
For the parameter choice $\Delta = \delta = G$, $\ket{011}$ is resonantly excited to $\ket{\phi_{f,-}}$. 
Then the state in Eq. \eqref{EqPsi1} is excited to
\begin{align*}
\ket{\phi_{1}} = c_0 \ket{\phi_{e,-}} + c_1 \ket{\phi_{f,-}}.
\end{align*}
From here, excitation exchange to the ancillary system $c_1$ by $H_{\rm anc}$ transfers the system to $\ket{\psi}\ket{1}_{c_1}$. Cooling of the ancillary system by $L_{c_j}$ recovers the original state $\ket{\psi}$. The last two steps can be described as an effective decay process from $\ket{\phi_{1}}$ to $\ket{\psi}$ (cf. Eq. \eqref{EqEngDecay}), which, together with the coherent drive from $\ket{\phi_1}$ to $\ket{\psi_1}$, realizes the desired error correction on qubit $1$. Errors on other qubits are corrected in an analogous fashion, utilizing the ancillary systems $c_j$.
We will later verify that this procedure indeed realizes the desired operators \eqref{EqLcorr}. We remark that the correction of several types of errors can also be realized sequentially rather than simultaneously. In this case only a single ancillary system $c$ is required. For correcting local bit flips, individual time slots $T_1$, $T_2$, $T_3$ can be assigned for correcting bit flip errors on qubit 1, qubit 2, and qubit 3 such that a sequence repeating these dedicated time slots $T_1,T_2,T_3,T_1,T_2,T_3,T_1...$ corrects the errors on the individual qubits one after the other.

Apart from the error-correcting mechanism, the scheme also entails undesired processes where spins are not flipped in accordance with the majority vote. As illustrated in Fig.~\ref{FigErrors}, this includes processes such as $\ket{100}\rightarrow\ket{110}$ or $\ket{000}\rightarrow\ket{100}$. For a suitable choice of parameters, these intrinsic errors can be strongly suppressed.
For example, the undesired excitations from $\ket{100}$ to the f-excited state $\ket{\chi_f} = (\ket{1f0}+\ket{10f})/\sqrt{2}$ and from $\ket{011}$ to the e-excited state $\ket{\chi_e} = (\ket{0e1}+\ket{01e})/\sqrt{2}$, shown in Fig. \ref{FigErrors} a), are suppressed by our parameter choice of $\Delta = \delta = G$. This can be understood as follows: $H_{\rm osc}$ couples $\ket{\chi_f}$ to $\ket{100}\ket{1}_b$. Due to constructive interference between the two terms in $\ket{\chi_f}$, this coupling has a strength of $\sqrt{2} G$.
The resulting dressed states thus reside at energies $\Delta_{2,\pm} \approx (1 \pm \sqrt{2}) G$ such that neither of the dressed states are in resonance with the drive. As a consequence, these processes taking the system away from the desired state will be much slower than the resonant processes correcting the errors. Excitation from $\ket{011}$ to $\ket{\chi_e}$ is suppressed for the same reason. The same arguments hold for states after single errors on other qubits.
The resulting slow process leads to a loss of population from the subspace of single-error states to the subspace of double-error states, which we discuss below.

In addition, the operation of the protocol also causes losses from the logical subspace, shown in Fig. \ref{FigErrors} b): $\ket{000}$ is excited to $\ket{\xi_{f}} = (\ket{f00} + \ket{0f0} + \ket{00f})/\sqrt{3}$ and $\ket{111}$ to $\ket{\xi_{e}} = (\ket{e11} + \ket{1e1} + \ket{11e})/\sqrt{3}$. These superposition states consist of three terms and couple to $\ket{000}\ket{1}_b$ and $\ket{111}\ket{1}_a$ with a coupling strength $\sqrt{3} G$. The energies of the resulting dressed states are therefore given by $\Delta_{3,\pm} = (1 \pm \sqrt{3}) G$. Hence, this excitation is also off-resonant with respect to the drive, leading only to a weak additional error process which is corrected by the action of the scheme.
As a result, the restoration of the memory state $\ket{\psi}$ from states with a single error is enhanced, whereas losses from $\ket{\psi}$ to states with a single error $\ket{\psi_j}$, and from $\ket{\psi_j}$ to states with two errors $\ket{\psi_{jk}}$ are suppressed.
Below, we optimize the operation of the scheme by the available parameters.

For our scheme, the precise parameters of the couplings $\Omega$, $G$, $g$, and $\kappa$ are not critical. To maintain the coherence of the codeword $\ket{\psi}=c_0\ket{000}+c_1\ket{111}$ under the correction in Eq. \eqref{EqLcorr}, we require that the rate $\Gamma_{\rm corr}$ is the same for both parts of the superposition, which is for example fulfilled if $\Omega$, $G$, and $g$ are identical for both paths illustrated in Fig.~\ref{FigMechanism}. To achieve a maximum correction rate, the detunings $\Delta$ and $\delta$ need to be tuned with an accuracy of at least $\kappa_{\rm eng}$.

\subsection{Correction of other types of errors}

Our scheme can be generalized to correct correlated bit flip or phase errors, as detailed in App. \ref{AppOtherErrors}.

The action of correlated spin flips, $L_{X} = L_{x_1} + L_{x_2} + L_{x_3} = \sqrt{\Gamma_X} (\sigma^x_1 + \sigma^x_2 + \sigma^x_3$) leaves the system in a superposition state of all single-qubit bit-flip errors, $\ket{\psi_X} = (\ket{\psi_1} + \ket{\psi_2} + \ket{\psi_3}) / \sqrt{N}$. By replacing the coupling of the ions to the three ancillary modes in equation \eqref{EqHanc} by a coherent coupling to a single ancillary mode, $c_j \rightarrow c$, 
\begin{align*}
H_{\rm anc} = g c^\dagger \sum_{j=1}^3 e^{i \delta_c t} \left( \ket{0}_j \bra{e} + \ket{1}_j \bra{f} \right) + H.c.
\end{align*}
The drive in equation \eqref{EqHdrive} also coherently acts on the three qubits. Correlated errors of all qubits are thus coherently mapped on $c$. Dissipating the excitations of $c$ by $L_c = \sqrt{\kappa} c$ (cf. equation \eqref{EqDecay} then realizes the single collective jump operator $L^{X}_{\rm corr} = L^{x_1}_{\rm corr} + L^{x_2}_{\rm corr} + L^{x_3}_{\rm corr}$.

To generalize the scheme to phase flips, we perform the mapping $\ket{0} \rightarrow \ket{+}$ and $\ket{1} \rightarrow \ket{-}$, where $\ket{\pm} = (\ket{0}\pm\ket{1})/\sqrt{2}$. The resulting codeword is given by $\ket{\psi}=c_+ \ket{+++} + c_- \ket{---}$. Phase flips on the second qubit are corrected by the action of the jump operator
\begin{align*}
L^{z_2}_{\rm corr} = \sqrt{\Gamma_{\rm corr}} \sigma^z_2 \frac{\mathbbm{1} - \sigma^x_{1} \sigma^x_2}{2} \frac{\mathbbm{1} - \sigma^x_2 \sigma^x_{3}}{2}.
\end{align*}
Errors on the first and third qubit are corrected analogously. To correct for correlated phase flips, we proceed as outlined above and replace the couplings of the qubits to individual ancillary modes $c_j$ by a coherent coupling of all three qubits to a single mode $c$. This yields the collective jump operator $L^{Z}_{\rm corr} = L^{z_1}_{\rm corr} + L^{z_2}_{\rm corr} + L^{z_3}_{\rm corr}$.

\subsection{Effective model}
In order to analytically verify that our scheme realizes the desired operators, we use the effective operator formalism \cite{EO} to adiabatically eliminate the excited degrees of freedom, which are coupled to the stable ground states by perturbative coherent couplings and are only weakly populated. As shown in App.~\ref{AppAnalysis}, the adiabatic elimination of the internal levels $\ket{e}$ and $\ket{f}$ leads to effective ground state dynamics (involving only the internal levels $\ket{0}$ and $\ket{1}$ with motional modes in the ground state) that is described by an effective master equation with jump operators
\begin{align}
L_{{\rm eff},j}
&= \sum_{n=1}^3 \sqrt{\kappa_{{\rm eff},n}} \left( \sigma^-_j P_{n_1=n} + \sigma^+_j P_{n_0=n} \right).
\label{EqLeff}
\end{align}
These jump operators contain the desired error-correcting terms ($n=1$) given by Eq.~\eqref{EqLcorr}, which result in a decay of the subspace of single-error states to the logical manifold at a large effective rate $\Gamma_{\rm corr} \equiv \kappa_{{\rm eff},1} \approx \Omega^2 / \kappa_{\rm eng}$, where $\kappa_{\rm eng}=g^2/\kappa$ is the engineered cooling rate introduced in Eq.~(\ref{EqEngDecay}).
In addition to these error-correcting processes, we also obtain weak undesired decay terms (see Fig.~\ref{FigErrors}). These include (i) terms with $n=3$ that act on the logical manifold introducing single-qubit errors (e.g. $\ket{000}\rightarrow\ket{100}$) at a rate $\kappa_{{\rm eff},3}$ and (ii) terms  with $n=2$, transferring single-error states to uncorrectable double-error states (e.g. $\ket{110}\rightarrow\ket{100}$).
As discussed above these processes are detuned by an amount $\sim G$ and are therefore strongly suppressed. We verify this in App.~\ref{SecEffective}.

\begin{figure}[t]
\centering
\includegraphics[width=8.6cm]{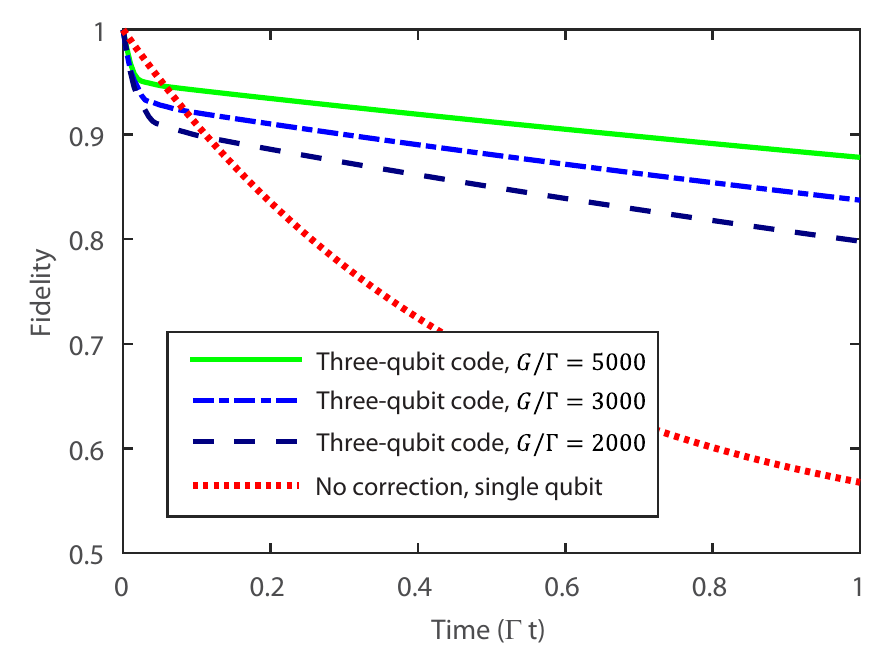}
\caption{
Autonomous three-qubit error correction of spin flip noise compared to single-qubit decay. We simulate the dynamics of a system of three physical qubits subjected to individual spin flip errors at a rate $\Gamma$ under the action of the error correction scheme by calculating of the full master equation $\rho(t)$ restricted to at most one excitation. The shown results 
are obtained from optimizing the fidelity (see main text) at $t=1/\Gamma$ by the choice of $\kappa_{\rm eng}$ and $\Omega$ for each considered value of the sideband coupling  $G$. For comparison, the red dotted line displays the decay of a single qubit subject to spin flips at a rate $\Gamma$. We find that the implementation of our scheme in trapped ions leads to a significantly reduced decay: Assuming a sideband coupling of $G = 5000 ~ \Gamma$ (green solid line), a fidelity of $F \approx 0.9$ of the dissipatively protected logical qubit (encoded in three decaying physical qubits) is maintained after a time $t \sim 1/\Gamma$.
}
\label{FigCurves}
\end{figure}

\section{Performance of the scheme}

We analyze the dynamics of our protocol analytically and numerically. For the former, we use the error correction rate $\Gamma_{\rm{corr}}=\kappa_{{\rm eff},1} $ and the rates of undesired processes $\kappa_{{\rm eff},2} $ and $\kappa_{{\rm eff},3} $ derived in App.~\ref{SecEffective} to describe the effective dynamics of the scheme by a system of coupled rate equations (see App.~\ref{SecRate}).
For the numerical analysis, we simulate the full dynamics of the system before using the effective operator formalism.
Here the dynamics of the system is modeled by a master equation $\dot{\rho}=\mathcal{L}_{\rm total}(\rho)$ with the Liouvillian $\mathcal{L}_{\rm total} = \mathcal{L}_{\rm noise} + \mathcal{L}_{\rm corr}$, where $\mathcal{L}_{\rm noise}$ describes the noise processes we aim to correct for (see Eq. \eqref{EqLiouvillianNoise}), and $\mathcal{L}_{\rm corr}$ contains the physical couplings required for the error correction scheme,
\begin{align}
\mathcal{L}_{\rm corr}(\rho) = -i [H_{\rm total, eng},\rho] + \sum_{j=1}^3 \mathcal{D}[L_{{\rm eng},j}](\rho).
\label{EqLiouvillian}
\end{align}
with the Hamiltonian $H_{\rm total, eng}$ given in Eq.~(\ref{EqHtot}) and the jump operators $L_{{\rm eng},j}$ given in Eq.~(\ref{EqEngDecay}).
The error-correcting code is assumed to operate on three physical qubits that are each subject to spin flip errors acting at a rate $\Gamma$. We calculate the fidelity $F(t)={\rm Tr}\left\{\rho(t)|\psi(0)\rangle\langle\psi(0)|\right\}$ with respect to the initial state $|\psi(0)\rangle=(\ket{000}_L+i\ket{111})/\sqrt{2}$ and compare the result with the decay of a single qubit subject to spin flips (cf. Eq. \eqref{EqSpinFlip}), that is initially prepared in the state $\ket{\psi(0)}=(\ket{0}+i\ket{1})/\sqrt{2}$.
For different sideband coupling strengths, we numerically optimize the parameters $\kappa_{\rm eng}$ and $\Omega$ to achieve maximum fidelity at the time $t = 1 / \Gamma$. The number of excitations in the simulation is limited to at most one. Given that we mostly operate in a regime of weak driving $\Omega^2 \ll \kappa_{\rm eng}^2, G^2$ this constitutes a good approximation. Truncating at higher numbers of excitations is found not to result in a notable difference of the evolution.

The results are shown in Fig.~\ref{FigCurves}, where we plot the dynamical evolution of the system for different values of $G/\Gamma$ (see Fig.~\ref{FigCurvesApp} in App. \ref{AppAnalysis} for a wider range of sideband couplings).
It can be seen that applying the dissipative three-qubit error correction code yields a significant advantage compared to using a single decaying qubit. For $G = 5000 ~ \Gamma$ the code maintains a fidelity close to $0.9$ at $t \sim 1/\Gamma$, where the single-qubit fidelity has almost dropped to the steady state value of $0.5$.

The fidelity decays in the form of a hockey-stick-shaped curve (compare Fig.~\ref{FigCurves}). As can be shown by means of the rate equation model introduced in App.~\ref{SecRate}, it features a fast initial drop to a fidelity $F_0 = 1 - E_0$ with $E_0 \approx \Gamma_{\rm err} / \Gamma_{\rm corr}$ which is characteristic for time-continuous quantum error-correcting schemes.
Here, $\Gamma_{\rm err} = 3(\Gamma + \kappa_{\rm eff, 3})$ is the total rate at which processes transfer states out of the logical subspace (see App. \ref{SecRate} for details). The drop is due to the fact that the loss from the codespace is proportional to its population and thus sets in immediately, whereas the correction processes only become effective once the correctable single-error states $\ket{\psi_j}$ become populated.
The inital drop is followed by an exponential decay with an effective decay rate $\Gamma_{\rm eff} \approx \Gamma_{\rm err} \Gamma_{\rm leak} / \Gamma_{\rm corr}$, where $\Gamma_{\rm leak} = 2(\Gamma + \kappa_{\rm eff, 2})$ is the total leakage rate from the subspace of single-error states to the subspace of two-error states (App. \ref{SecRate}). As is apparent from Fig.~\ref{FigCurves}, the effective decay rate in the presence of error correction is substantially reduced as compared to the single-qubit decay rate.
Assuming perfect error correction ($G \rightarrow \infty$), the detrimental rates approach the bare spin flip rates, thus $\Gamma_{\rm err} \rightarrow 3 ~ \Gamma$, $\Gamma_{\rm leak} \rightarrow 2 ~ \Gamma$, and the initial drop and effective decay rate become $E_0 \approx \Gamma / \Gamma_{\rm corr}$ and $\Gamma_{\rm eff} \approx 6 ~ \Gamma^2 / \Gamma_{\rm corr}$, as expected for a continuous implementation of the three-qubit code~\cite{Ippoliti2014,Paz1998}.\\

For finite sideband coupling $G$, we maximize the fidelity for long times by minimizing $\Gamma_{\rm eff}$ (the details of the optimization are given in App. \ref{SecMinimization}). This leads to an optimal parameter choice $\Omega \approx \kappa_{\rm eng} \approx 1.2 \sqrt[3]{\Gamma G^2}$, which allows for a correction rate $\Gamma_{\rm corr} \approx \kappa_{\rm eng} / 3$ and error / leakage rates $\Gamma_{\rm err} \approx \Gamma_{\rm leak} \approx 3 ~ \Gamma$. The resulting initial drop and effective decay rate are $E_0 \approx 10 (\Gamma/G)^{2/3}$ and $\Gamma_{\rm eff} \approx 27 (\Gamma / G)^{2/3} \Gamma$. Alternatively, the protocol can be optimized for short operation times by minimizing the initial drop $E_0$.

\begin{figure}[t]
\centering
\includegraphics[width=8.6cm]{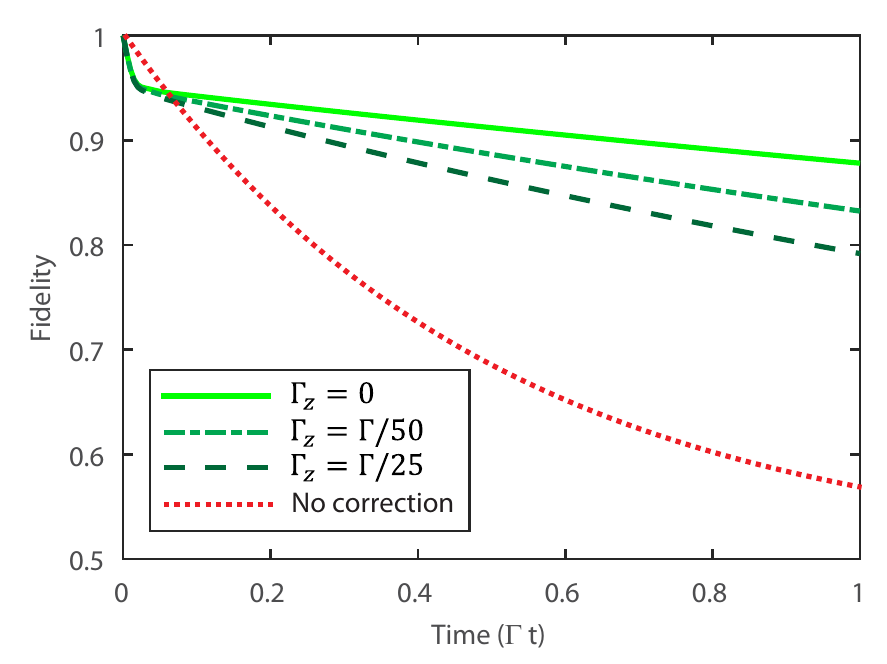}
\caption{
Dissipative correction of spin flips in the presence of dephasing. The plot shows the decrease in fidelity (compare Fig.~\ref{FigCurves}) for three different values of the ratio $\Gamma_z/\Gamma$, where $\Gamma_z$ ($\Gamma$) is the rate at which an individual phase flip (individual bit flip) occurs, assuming a sideband coupling $G=5000 ~ \Gamma$: $\Gamma_z=0$ (light green solid), $\Gamma_z=1/25 ~ \Gamma$ (long green dash), and $\Gamma_z=1/50 ~ \Gamma$ (short dark green dash). As reference, the plot also includes the decay of a single uncorrected ion under bit flips at a rate $\Gamma$ with $\Gamma_z=0$ (red dashed line).
}
\label{FigImperfections}
\end{figure}

\subsection{External Imperfections}

In the following, we discuss imperfections that can occur in realistic setups and may lead to a reduction of the fidelity of the logical state.

Our scheme is designed to correct for one type of individual or collective errors (for example for correcting either bit-flip or phase errors). Errors other than the targeted type cannot be corrected simultaneously in the present version of the scheme. The performance of a scheme that corrects bit flips ($\sigma^x$-errors), will for example be degraded by the presence of ``complementary'' $\sigma^z$-errors,

\begin{align*}
L^{z_j} &= \sqrt{\Gamma_z} \sigma^z_j
\\
L^{Z} &= \sqrt{\Gamma_Z} \sum_{j=1}^N \sigma^z_j.
\end{align*}
These errors contribute to the effective error-correcting dynamics and cause a decaying envelope for the population of the codeword $\ket{\psi}$, $F_{\rm comp}(t) = F(t) e^{-(3\Gamma_z+\Gamma_Z) t}$.
To limit the additional error to the few-percent level at $t \sim 1/\Gamma$, it is thus required that $\Gamma_k \lesssim \Gamma / 50 - \Gamma / 100$ depending on the kind of error process $k$. In Fig.~\ref{FigImperfections}, we plot the evolution for $G/\Gamma = 5000$ and individual phase flips with $\Gamma_z/\Gamma=0,1/50,1/25$, finding that $\Gamma_z/\Gamma = 1/50$ leads to a decrease in fidelity of about $\sim 0.05$ at $t \sim 1/\Gamma$. The decrease is less pronounced for smaller $G/\Gamma$.

Next, we address the effect of decoherence associated with the decay of the excited degrees of freedom. More specifically, we include the decay of the excited states $\ket{e}$ and $\ket{f}$ by spontaneous emission described by the jump operators $L_{mn,j} = \sqrt{\gamma_{m,n}} \ket{m}_j \bra{n}$, where $m \in \{e,f\}$ and $n \in \{0,1\}$, in the master equation. Also the oscillator modes are assumed to undergo decay $L_{r} = \sqrt{\kappa_r} r$, where $r \in \{a,b\}$. For simplicity we assume that $\gamma_{m0} = \gamma_{m1} = \gamma_m/2$ (for $m \in \{e,f\}$), $\gamma_e = \gamma_f = \gamma$ and $\kappa_{b} = \kappa_{b}$.
These imperfections do not change the effective couplings significantly. Analytically, they can easily be taken into account as imaginary parts in the detunings \cite{EO}: $\tilde{\Delta} \rightarrow \tilde{\Delta} - i \gamma_m/2$ and $\tilde{\delta} \rightarrow \tilde{\delta} - i \kappa_m/2$. For $\kappa_{\rm eng} \gg \gamma_m, \kappa_m$ we can safely assume that the rates $\kappa_{{\rm eff},1-3}$ are not affected.
Numerical simulations show that for the parameters used in Fig.~\ref{FigCurves}, $\{G/\gamma_m, G/\kappa_m\} \sim 10^3$ or $\{\gamma_m, \kappa_m\} \sim \Gamma$, respectively, can be tolerated. Note that in the quantum jump formalism, dephasing of states $\ket{e}$ and $\ket{f}$ enter into the effective non-Hermitian Hamiltonian in a similar manner as the decay. Decay and dephasing thus only differ in their dynamics after a decay causing an error; we therefore expect similar results for dephasing.

\begin{figure}[t]
\centering
\includegraphics[width=8.6cm]{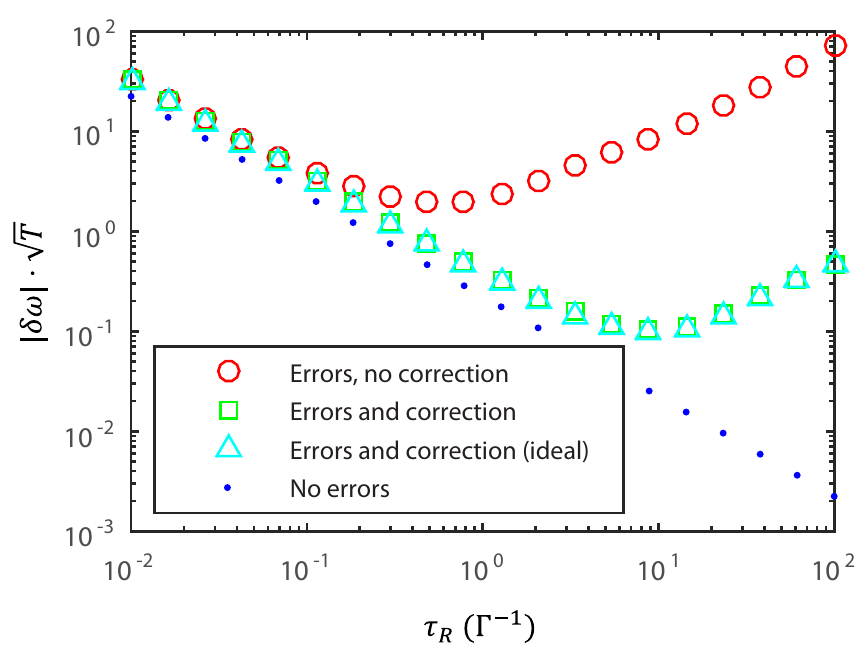}
\caption{
Quantum error correction enhanced sensing. The plot shows the normalized measurement precision $|\delta\omega| \sqrt{T}$ versus the Ramsey time $\tau_{\rm R}$ for a Ramsey measurement subject to spin flips. The action of the error-correcting scheme can be seen to improve the achievable sensitivity by about an order of magnitude: The shown results display the normalized measurement precision assuming errors and (i) no error correction, (ii) error correction using the ideal jump operators given in Eq.~(\ref{EqLcorr}), and (iii) error correction including undesired processes using the jump operators given in Eq.~(\ref{EqLeff}). As reference, the precision in the absence of errors (iv) is also shown. For the numerical simulation, we use $G = 5000 ~ \Gamma$, where $\Gamma$ is the bit flip rate, and the optimized parameters $\Omega = 4 \kappa_{\rm eng}/5$ and $\kappa_{\rm eng} = 1.2 \sqrt[3]{\Gamma G^2}$ discussed in the text, and restrict the full master equation to the subspace of at most one excitation. 
}
\label{FigSensitivity}
\end{figure}

\section{Application to quantum metrology}

Quantum error correction protocols find attractive applications for quantum metrology, as has been proposed for example in~\cite{Dur,Arrad,Kessler}. Following these ideas, we explore the application of our error correction scheme for improving high-precision measurements of weak magnetic fields with ions. To this end, we study a prototypical setting, where a Ramsey-type protocol (see Fig. \ref{FigMetrology}a) is used to measure a signal originating from a field in the $z$-direction acting on several probe particles $H = (\omega/2) \sum_{j=1}^N \sigma^z_j$. We commence by considering three probe particles constituting one logical qubit. The measurement sequence involves four steps. (i): Starting from the initial state $\ket{000}$, a first Ramsey pulse (a $\pi/2$ rotation on the logical qubit) prepares the superposition state $\ket{\psi(0)} = (\ket{000} + \ket{111})/\sqrt{2}$. (ii) During a Ramsey waiting time of duration $\tau_{\rm{R}}$, the superposition state picks up a relative phase $\phi(\tau_{\rm{R}})=3\omega \tau_{\rm{R}}$, such that $\ket{\psi(\tau_{\rm{R}})} = (\ket{000} + e^{-i \phi(\tau_{\rm{R}})} \ket{111})/\sqrt{2}$. (iii) A second Ramsey pulse (another $\pi/2$ rotation on the logical qubit) transforms the evolved state into $\ket{\psi'(\tau_{\rm{R}})} =\cos(\phi(\tau_{\rm{R}})/2)\ket{111}+i\sin(\phi(\tau_{\rm{R}})/2)\ket{000}$. (iv) Finally, a measurement in the $\sigma^z$-basis is performed on one of the qubits. The probability to detect the first (or any other) qubit in state $\ket{1}$ is given by $P_1=\cos^2(\phi(\tau_{\rm{R}})/2)$, which allows one to infer the phase $\phi(\tau_{\rm{R}})$ and thus the signal strength $\omega$. As explained in App. \ref{App_Metrology_Sensitivity}, the sensitivity of this measurement is given by $|\delta \omega|=1/(N \sqrt{\tau_{\rm{R}}}\sqrt{T})$, where $N$ is the number of probe particles ($N=3$ in our case) and $T$ is the total measurement time ($T=m \tau_{\rm{R}}$, where $m$ is the number of runs). Increasing the Ramsey time leads to an improvement of the measurement precision (see Fig.~\ref{FigSensitivity}).

In the presence of noise, the precision of the measurement changes significantly~\cite{Acin,Huelga}, given that both the amplitude and period of the fringes (cf. Fig. \ref{FigMetrology}) are affected by the noise. Assuming that the aim of the Ramsey protocol is to detect a signal in z-direction, the detrimental effect of noise in the transverse direction can be mitigated by the error correction protocol. Noise processes acting along the same direction as the signal cannot be distinguished from the signal and thus cannot be targeted by error correction; we consider their effect in App.~\ref{App_Metrology_Scalability}.
In the Ramsey-type sensing scheme above, we consider transverse noise in the form of local spin flips, $L_{x_j}=\sqrt{\Gamma}\sigma_j^x$. This degrades the state with a resulting decrease of the fringe contrast (as can be seen in Fig.~\ref{FigMetrology}).
As shown in Fig.~\ref{FigSensitivity}, the normalized sensitivity $\sqrt{T}|\delta \omega|$ improves up to Ramsey waiting times $\tau_R \sim \Gamma^{-1}$, which limits the achievable measurement precision. The application of our quantum error correction scheme during step (ii) of the protocol (free evolution for a time $\tau_{\rm{R}}$) thus reduces the speed at which the state is degraded. This shifts the optimum Ramsey waiting time to higher values, which improves the achievable measurement sensitivity. Fig.~\ref{FigSensitivity} illustrates this effect for realistic experimental parameters and demonstrates that the presented error correction scheme significantly improves the performance of the Ramsey sequence compared to the situation where transversal errors are not corrected. The achievable sensitivity in the presence of both transversal and parallel noise ($\sigma^x$ and $\sigma^z$ errors) is shown in App.~\ref{App_Metrology_Scalability}. 

So far our discussion involves one logical qubit. The scheme can be scaled up to involve $N_{\rm log}$ logical qubits using a segmented trap~\cite{SegmentedTrap1,SegmentedTrap2}, as shown in Fig. \ref{FigConcepts} c) and explained in App.~\ref{App_Metrology_Scalability}. In this case, the potential barriers can be ramped up that divide the trap into segments confining three ions each. For sufficiently high barriers, the motional modes of the ion triples are independent from each other such that each logical qubit can be individually protected.
This ramping into individual segments can be performed after the qubits are merged into an entangled state containing all qubits. Thereby the protocol can be used to obtain maximal quantum enhanced sensitivity while simultaneously being protected against spin flip errors.

\section{Conclusions and Outlook}\label{SecConclusions}

In summary, we have presented an autonomous quantum error correction scheme for trapped ions that stabilizes an encoded qubit by coupling it to a engineered reservoir. This protocol does not require the use of a classical measurement apparatus or classical feedback loops, since the error correction mechanism results from a built-in back-action mechanism induced by engineered dissipative processes.
From a practical perspective, avoiding the measurement procedure is a major advantage since measurements typically require scattering thousands of photons, which is a slow process and heats up the motional mode.
While our scheme relies on the coupling of internal degrees of freedom to motional modes that underpins the realization of quantum gates in trapped ions \cite{CiracZoller, ExperimentWineland, ExperimentBlatt, Monz}, it does not involve a sequence of gates, as in standard approaches to quantum error correction. Instead, our protocol uses the interaction between internal degrees of freedom and motional modes directly and requires only always-on couplings that act simultaneously and continuously on the encoded qubit. In this way, the role of ancillary systems to which the error syndromes are mapped, can be naturally played by the motional degrees of freedom, which allows one to continuously remove errors by means of standard sympathetic cooling of the motional modes.

The proposed error correction scheme can for example be used for quantum sensing, as we showed for the case of one logical qubit. By protecting random superposition states of three qubits we significantly increase the lifetime of the coherent oscillations in a Ramsey measurement. As a result, the sensitivity can be improved and the optimal measurement time can be shifted to higher values. For scaling up this approach, segmented ion traps can be used that can accommodate several logical qubits in different trap segments.

Our protocol pushes forward the boundary of dissipation-driven quantum information processing towards universal dissipative quantum computing. Engineered dissipation has already found useful applications for quantum state preparation, where the initial state is destroyed in the process of preparing a desired target state. In contrast, quantum error-correcting schemes need to preserve coherences in the initial state of a quantum system. Since such quantum error correction inherently relies on dissipation to get rid of entropy, it naturally fits into the framework of engineered dissipation. The realization of dissipative protocols that are capable of manipulating non-orthogonal quantum states while maintaining their coherence is an essential step in the endeavor to perform autonomously stabilized quantum information processing tasks.

The principles we have employed to implement a dissipative three-qubit code code can be taken further towards more sophisticated quantum error-correcting codes. Here it will be interesting to tailor suitable many-body dissipation to protect larger classes of stabilizers with the aim to realize, for example, topological error correction. As the couplings assumed for the implementation of our scheme are generic couplings of a register of qubits to oscillators/ancilla qubits, the presented working mechanism can be adopted to other physical systems such as superconducting architectures and Rydberg atoms.

We remark that the dissipative confinement of a quantum system to a desired subspace could also be useful for quantum simulations of lattice gauge theories, where devising methods for limiting the dynamics of a system to the gauge-invariant part of the Hilbert space is a key challenge for the development of future quantum simulators~\cite{WieseReview,ExpLGTPaper}.
From a broader perspective, the design of dissipative maps can find a wide range of applications for quantum information processing including dissipative schemes for entanglement distillation~\cite{DissDistillation}, generalized quantum measurements and the simulation of exotic quantum channels~\cite{ChannelConstruction}.

\section*{Acknowledgments}

We gratefully acknowledge discussions with Philipp Schindler, Esteban Martinez, Thomas Monz, Martin van Mourik, Rainer Blatt, and Wolfgang D\"ur. The research leading to these results has received funding from the Army Research Laboratory Center for  Distributed  Quantum Information via the project SciNet, the
European Research Council under the European Union's Seventh Framework Programme (FP/2007-2013), the ERC Grant Agreement n. 306576, the Villum Kann Rasmussen Foundation, and the EU project SIQS. FR acknowledges support by a Feodor-Lynen fellowship from the Humboldt Foundation.

\appendix

\section{Correction of other types of errors}

\label{AppOtherErrors}

In the following we discuss the generalization of our scheme to other types of errors. We start by discussing correlated spin flips. Then we generalize the scheme to individual phase flips, and finally explain how to correct collective dephasing.

\subsection{Correlated spin flips}

\label{SecCorrelated}

The correction of the error syndromes $\sigma^x_1$, $\sigma^x_2$, and $\sigma^x_3$ we have discussed so far does not interfere with each other and is performed using separate cooling resources. 
Instead of targeting individual errors, our scheme can as well be formulated to correct correlated errors, such as, e.g., collective spin flips,
\begin{align*}
L_{X}
&= L_{x_1} + L_{x_2} + L_{x_3} = \sqrt{\Gamma_X} (\sigma^x_1 + \sigma^x_2 + \sigma^x_3).
\end{align*}
The action of such an error leaves the system in the state
\begin{align*}
\ket{\psi_X} = \frac{1}{\sqrt{3}} (\ket{\psi_1} + \ket{\psi_2} + \ket{\psi_3}) .
\end{align*}
$\ket{\psi}$ can be restored by the single-qubit correcting operators in equations \eqref{EqLcorrX2}--\eqref{EqLcorrX3}. In this case, however, a simpler strategy is to engineer a single correcting operator
\begin{align*}
L^{X}_{\rm corr} 
&= L^{x_1}_{\rm corr} + L^{x_2}_{\rm corr} + L^{x_3}_{\rm corr},
\end{align*}
which is given by the sum of the operators in equations \eqref{EqLcorrX2}--\eqref{EqLcorrX3}.
We implement this interaction by replacing the individual mapping from single ions onto single ancilla systems in equation \eqref{EqHanc} by a collective mapping of all ions onto a single ancillary system $c$, using the coupling
\begin{align}
H_{\rm anc} = g e^{i \delta_c t} c^\dagger \sum_{j=1}^3 \left( \ket{0}_j \bra{e} + \ket{1}_j \bra{f} \right) + H.c.
\label{EqHanc2}
\end{align}
In the presence of only correlate noise, this approach is preferable since it reduces the required resources compared to realizing three operators. If both types, uncorrelated and correlated errors are present, the operations in equations \eqref{EqLcorrX2}--\eqref{EqLcorrX3} for the correction of uncorrelated errors suffice for the correction of both.

\subsection{Phase errors and collective dephasing}

Instead of correcting for spin errors, our scheme can as well be formulated to correct phase errors
\begin{align*}
L_{z_j} &= \sqrt{\Gamma_z} \sigma^z_j,
\end{align*}
where $\sigma^z_j = \ket{0}_j \bra{0} - \ket{1}_j \bra{1}$ applies a phase change to qubit $j$.
To correct such errors we make the replacements $\ket{0} \rightarrow \ket{+} = (\ket{0}+\ket{1})/\sqrt{2}$, $\ket{1} \rightarrow \ket{-} = (\ket{0}-\ket{1})/\sqrt{2}$ throughout the protocol. Our codeword is then
\begin{align*}
\ket{\psi_z} = c_+ \ket{+++} + c_- \ket{---},
\end{align*}
and the states after an individual phase flip are given by
\begin{align*}
\ket{\psi^z_j} = \sigma^z_j \ket{\psi_z},
\end{align*}
e.g., $\ket{\psi^z_2} = c_+ \ket{+-+} + c_- \ket{-+-}$ for a phase flip on the second of three qubits. Decay from $\ket{+++}$ and from $\ket{---}$ is again well-distinguishable.
The codeword $\ket{\psi_z}$ is recovered by the jump operators
\begin{align}
L^{z_1}_{\rm corr} = \sqrt{\Gamma_{\rm corr}} \sigma^z_1 \frac{\mathbbm{1} - \sigma^x_1 \sigma^x_2}{2} \frac{\mathbbm{1} - \sigma^x_1 \sigma^x_3}{2}
\label{EqLcorrZ1}
\\
L^{z_2}_{\rm corr} = \sqrt{\Gamma_{\rm corr}} \sigma^z_2 \frac{\mathbbm{1} - \sigma^x_1 \sigma^x_2}{2} \frac{\mathbbm{1} - \sigma^x_2 \sigma^x_3}{2}
\label{EqLcorrZ2}
\\
L^{z_3}_{\rm corr} = \sqrt{\Gamma_{\rm corr}} \sigma^z_3 \frac{\mathbbm{1} - \sigma^x_1 \sigma^x_3}{2} \frac{\mathbbm{1} - \sigma^x_2 \sigma^x_3}{2}
\label{EqLcorrZ3}
\end{align}
with a correction rate $\Gamma_{\rm corr}$ for every single error.
The operators are identical to the ones in equations \eqref{EqLcorrX2}--\eqref{EqLcorrX3} when making the above replacements.

Another experimentally common error is correlated phase noise, or, collective dephasing \cite{Monz}. It is described by a single jump operator
\begin{align}
L_{Z} &= L_{z_1} + L_{z_2} + L_{z_3} = \sqrt{\Gamma_Z} (\sigma^z_1 + \sigma^z_2 + \sigma^z_3),
\label{EqLZCollective}
\end{align}
which causes a decay from $\ket{\psi_z}$ to the single-error state
\begin{align*}
\ket{\psi_Z} = \frac{1}{\sqrt{3}} (\ket{\psi^z_1} + \ket{\psi^z_2} + \ket{\psi^z_3})
.
\end{align*}
To correct for this error, we use a coupling to an single ancillary system,
\begin{align}
H_{\rm anc} = g e^{i \delta_c t} c^\dagger \sum_{j=1}^3 \left( \ket{+}_j \bra{e} + \ket{-}_j \bra{f} \right) + H.c.
\label{EqHanc3}
\end{align}
This realizes the jump operator
\begin{align}
L^{Z}_{\rm corr} 
&= L^{z_1}_{\rm corr} + L^{z_2}_{\rm corr} + L^{z_3}_{\rm corr},
\label{EqLCollDephasing}
\end{align}
with the single-qubit correcting operators in equations \eqref{EqLcorrZ1}--\eqref{EqLcorrZ3}.

\section{Analysis}

\label{AppAnalysis}

In this appendix we provide a detailed analysis of the presented scheme. We derive analytical expressions for the effective rates in the system and provide a numerical comparison to the full dynamics. Considering a simplified rate equation model we assess the performance of the scheme and determine the optimal parameters of the scheme analytically.

\subsection{Effective open system dynamics}

\label{SecEffective}

We reduce the full dynamics of the model to effective dynamics of the ground states by adiabatically eliminating the excited degrees of freedom. To this end, we use the effective operator formalism \cite{EO}. Here we assume that the stable ground states are coupled to the excited states by perturbative coherent couplings $V=V_+ + V_-$, where $V_+$ ($V_-$) denotes (de-)excitation. The coherent and dissipative dynamics of the excited states is described by a non-Hermitian Hamiltonian
\begin{align*}
H_{\rm NH} = H_e - \frac{i}{2} \sum_k L_{k}^\dagger L_{k},
\end{align*}
where $H_e$ contains the coherent couplings between the excited states and $L_k$ are decay processes taking them to the ground states. Applying the formalism \cite{EO} we then obtain the effective operators
\begin{align*}
H_{{\rm eff}}
&= - \frac{1}{2} V_- H_{\rm NH}^{-1} V_+ + H.c.,
\\
L_{k, {\rm eff}}
&= L_{k} H_{\rm NH}^{-1} V_+.
\end{align*}
Here, $H_{\rm eff}$ is the effective Hamiltonian and $L_{k,{\rm eff}}$ are the effective jump operators. The resulting effective dynamics is described by an effective master equation
\begin{align}
\dot{\rho} = -i [H_{\rm eff}, \rho] + \sum_k \mathcal{D}[L_{k,{\rm eff}}](\rho).
\end{align}
From this reduced model we derive the rates for the error correction and leakage processes. 

Following the procedure of \cite{EO}, we first set up the non-Hermitian Hamiltonian $H_{\rm NH} = H_e - i L_{\rm eng}^\dagger L_{\rm eng} / 2$ describing the dynamics in the single-excitation manifold (higher excitations are neglected). It is of block-diagonal form
\begin{align*}
H_{\rm NH} = \sum_{\phi,f} H_{{\rm NH},\ket{\phi}\ket{1}_f}.
\end{align*}
Each of the blocks $H_{{\rm NH},\ket{\phi}\ket{1}_f}$ contains a single oscillator-excited state $\ket{\phi}\ket{1}_f$, where $\ket{\phi}$ can be any ground state of the system ions and $f \in \{a,b\}$ denotes the oscillator. For example,
\begin{align*}
H_{{\rm NH},\ket{100}\ket{1}_a} = & ~ \tilde{\Delta} \ket{e00} \ket{0}_a \bra{0} \bra{e00} + \tilde{\delta} \ket{100}\ket{1}_a \bra{1} \bra{100}
\nonumber
\\
&+ G (\ket{100}\ket{1}_a \bra{0} \bra{e00} + H.c.)
\end{align*}
is needed for the correction of a single error on qubit $1$,
\begin{align*}
H_{{\rm NH},\ket{100}\ket{1}_b} = & ~ \tilde{\Delta} \ket{\chi_f} \ket{0}_b \bra{0} \bra{\chi_f} + \tilde{\delta} \ket{100}\ket{1}_b \bra{1} \bra{100}
\nonumber
\\
&+ \sqrt{2} G \ket{100}\ket{1}_b \bra{0} \bra{\chi_f} + H.c.)
\end{align*}
participates in the undesired decay of $\ket{100}$, and 
\begin{align*}
H_{{\rm NH},\ket{111}\ket{1}_a} = & ~ \tilde{\Delta} \ket{\xi_e} \ket{0}_a \bra{0} \bra{\xi_e} + \tilde{\delta} \ket{111}\ket{1}_a \bra{1} \bra{111}
\nonumber
\\
&+ \sqrt{3} G \ket{111}\ket{1}_a \bra{0} \bra{\xi_e} + H.c.)
\end{align*}
mediates the undesired decay from $\ket{111}$.
The non-Hermitian Hamiltonian is formulated in terms of complex detunings $\tilde{\Delta} = \Delta - i \gamma/2$ and $\tilde{\delta} = \delta - i \kappa/2$, where $\gamma$ ($\kappa$) accounts for potential finite lifetime of the excited levels (motional modes used for interrogation). Further down, we will, however, make the assumption $\gamma, \kappa \ll \kappa_{\rm eng}$. Inversion of the blocks of the non-Hermitian Hamiltonian yields, e.g.,
\begin{align*}
\left(H_{{\rm NH},\ket{100}_a}\right)^{-1}
= & ~ \tilde{\Delta}_{1, \rm eff}^{-1} \ket{e00} \ket{0}_a \bra{0} \bra{e00}
\\
&+ \tilde{\delta}_{1, \rm eff}^{-1} \ket{100} \ket{1}_a \bra{1} \bra{100}
\nonumber
\\
&+ \tilde{G}_{1, \rm eff}^{-1} \left(\ket{100} \ket{1}_a \bra{0} \bra{e00} + H.c. \right)
\nonumber
\\
\left(H_{{\rm NH},\ket{100}\ket{1}_b}\right)^{-1}
= & ~ \tilde{\Delta}_{2, \rm eff} \ket{\chi_f} \ket{0}_a \bra{0} \bra{\chi_f}
\\
&+ \tilde{\delta}_{2, \rm eff} \ket{100}\ket{1}_b \bra{1} \bra{100}
\nonumber
\\
&+ \tilde{G}_{2,\rm eff} \ket{100}\ket{1}_b \bra{0} \bra{\chi_f} + H.c.)
\nonumber
\\
\left(H_{{\rm NH},\ket{111}_a}\right)^{-1}
= & ~ \tilde{\Delta}_{3, \rm eff}^{-1} \ket{\xi_e} \ket{0}_a \bra{0} \bra{\xi_e}
\\
&+ \tilde{\delta}_{3, \rm eff}^{-1} \ket{111} \ket{1}_a \bra{1} \bra{111}
\nonumber
\\
&+ \tilde{G}_{3, \rm eff}^{-1} \left(\ket{111} \ket{1}_a \bra{0} \bra{\xi_e} + H.c. \right)
\nonumber
\end{align*}
with effective detunings and couplings
\begin{align*}
\tilde{\Delta}_{n, \rm eff} &= \tilde{\Delta} - \frac{n G^2}{\tilde{\delta}},
\\
\tilde{\delta}_{n, \rm eff} &= \tilde{\delta} - \frac{n G^2}{\tilde{\Delta}},
\\
\tilde{G}_{n, \rm eff} &= \sqrt{n} G - \frac{\tilde{\Delta} \tilde{\delta}}{\sqrt{n} G}.
\end{align*}
By our choice of $\Delta = \delta = G$ and our assumption $|{\rm Im}(\tilde{\Delta})|,|{\rm Im}(\tilde{\delta})| \ll \kappa_{\rm eng}$ (where ${\rm Im}()$ denotes the imaginary part), we engineer the effective detuning involved in the correction mechanism to be small and only limited by the linewidth due to the engineered cooling,
\begin{align*}
\tilde{\Delta}_{1, \rm eff} \approx - \frac{i \kappa_{\rm eng}}{2}.
\end{align*}
In contrast, the terms with $n \neq 1$, i.e., $\tilde{\Delta}_{n,\rm eff} \approx (n-1) G$, are chosen to be large to render the undesired processes weak.
We identify the driving fields with the perturbative couplings between ground and excited subspaces, $H_{\rm drive} = V_+ + V_-$, where $V_+$ is the part responsible for the excitation out of the ground subspace. We obtain the effective jump operators \cite{EO}
\begin{align}
L_{{\rm eff},j}
&= L_{{\rm eng},j} H_{\rm NH}^{-1} V_+
\\
&= \sum_{n=1}^3 \sqrt{\kappa_{{\rm eff},n}} \left( \sigma^-_j P_{n_1=n} + \sigma^+_j P_{n_0=n} \right).
\label{EqLeffApp}
\end{align}
Here, $P_{n_{0}=n}$ ($P_{n_{1}=n}$) denote projectors onto all ground states with $n$ atoms in state $\ket{0}$ ($\ket{1}$). The decay rates of the effective processes are given by
\begin{align}
\kappa_{{\rm eff},n} &= \frac{\kappa_{\rm eng} \Omega^2}{4 |\tilde{\Delta}_{n,\rm eff}|^2}.
\label{Eqkappaeff}
\end{align}
For the parameters at hand we obtain
\begin{align}
\kappa_{{\rm eff},1} &\approx \frac{\Omega^2}{\kappa_{\rm eng}},
\\
\kappa_{{\rm eff},2} &\approx \frac{\kappa_{\rm eng} \Omega^2}{4 G^2},
\label{EqKappaEff2}
\\
\kappa_{{\rm eff},3} &\approx \frac{\kappa_{\rm eng} \Omega^2}{16 G^2}.
\label{EqKappaEff3}
\end{align}
Here, $\kappa_{{\rm eff},1}$ is the effective decay rate from the single-error states to the logical subspace containing the codeword and $\kappa_{{\rm eff},2/3}$ are intrinsic loss rates: $\kappa_{{\rm eff},2}$ is the leakage rate from the single-error states to the double-error states and $\kappa_{{\rm eff},3}$ is a decay rate from the logical states to the single-error states.
Comparing Eq. \eqref{EqLeffApp} with Eq. \eqref{EqLcorr}, thereby identifying $\Gamma_{\rm corr} = \kappa_{{\rm eff},1}$, we conclude that we have engineered the desired correction operators up to small terms $\mathcal{O}(\kappa_{\rm eng} \Omega^2/G^2)$ acting on manifolds other than the single-error subspace.
In addition to the effective decays considered so far, the effective operator formalism \cite{EO} also contains an effective Hamiltonian $H_{\rm eff}$, which in the case at hand is found to only contain minor Stark shifts.

\subsubsection{Strong driving effects}

It should be noted that the above expressions are obtained by means of a perturbative formalism, and are thus only correct for $\Omega \ll \kappa_{\rm eng}$. 
Numerical optimization shows, however, that the scheme is more effective for $\Omega \sim \kappa_{\rm eng}$ because this allows for faster correction of errors (cf. Ref. \cite{RRS}). We therefore include two strong driving effects: Power broadening and population of the excited states. Power broadening is relevant for the linewidth-limited process leading to $\kappa_{{\rm eff},1}$. Including it is easily achieved by replacing the bare linewidth of the excited states that mediate the correction of the single-error states, $\kappa_{\rm eng}$, by a broadened one, $\kappa_{\rm eng} + 2\Omega^2/\kappa_{\rm eng}$, which can be justified using adiabatic elimination \cite{RRS}. We find
\begin{align}
\kappa_{{\rm eff},1} 
= \frac{\kappa_{\rm eng} \Omega^2}{\kappa_{\rm eng}^2 + r \Omega^2}
\label{EqKappaEffStrong}
\end{align}
for the preparation rate. While considering a simple two-level model would yield $r = 2$ \cite{RRS}, the numerics in the case at hand turn out to be more accurate for $r \approx 2.5$. This can be understood from further excited levels coupled by the drive. The error rates in Eqs. \eqref{EqKappaEff2}--\eqref{EqKappaEff3} remain effectively unchanged by power broadening (under the reasonable assumption of $\Omega \ll G$).
Secondly, we need to take into account that due to the increased driving, part of the population resides in the excited state manifold, which we previously eliminated. The main contribution to this steady population comes from the off-resonant excitation of the codeword. The effect can be taken into account by replacing the fidelity of the codeword $F$ by $(1-f_e)F$, where the excited fraction $f_e$ is estimated to be \cite{RRS}:
\begin{align}
f_e \approx \frac{(\sqrt{3}\Omega)^2}{(\sqrt{3}-1)^2 G^2} + \frac{(\sqrt{3}\Omega)^2}{(\sqrt{3}+1)^2 G^2} \approx \frac{6 \Omega^2}{G^2}.
\label{EqPopulationStrong}
\end{align}
Here the expression $(\sqrt{3} \pm 1)^2 G^2$ has its origin in the energies $\pm \sqrt{3} G$ of the dressed states off-resonantly excited by the drives with detunings $G$.

\begin{figure}[t]
\centering
\includegraphics[width=8.6cm]{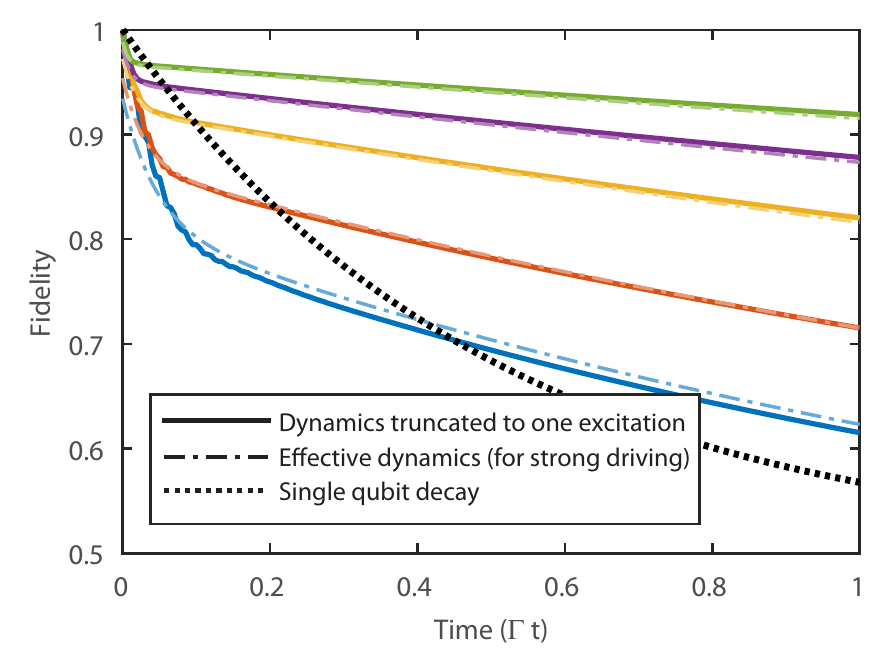}
\caption{
Comparison of the effective and full dynamics for the presented protocol. We simulate the dynamics of the system subjected to spin flip errors under the action of the error correction scheme for different ratios of the sideband coupling and the error rate, $G/\Gamma$ (from bottom to top: $G/\Gamma=500,1000,2500,5000,10000$). We plot results obtained by simulating the full master equation restricted to at most one excitation (solid lines), and the effective dynamics of the ground states with strong driving corrections (dash-dotted lines).
For each value of $G/\Gamma$, the fidelity at $t=1/\Gamma$ is optimized by the choice of $\kappa_{\rm eng}$ and $\Omega$. The results are compared with the corresponding decay of a single qubit (dotted line).
}
\label{FigCurvesApp}
\end{figure}

\subsubsection{Numerical results and comparison of the methods}

We simulate the evolution due to the effective operators in Eq. \eqref{EqLeffApp}, taking into account the strong driving effects in Eq. \eqref{EqKappaEffStrong} and Eq. \eqref{EqPopulationStrong}. The results after parameter optimization assuming sideband couplings $G/\Gamma = 500$ -- $10000$ are presented in Fig. \ref{FigCurvesApp}. We plot them together with the result from a simulation of the full master equation in Eq. \eqref{EqLiouvillian} truncated to a single excitation. We find that the effective dynamics agrees well with the full dynamics.

It can be seen that for $G/\Gamma = 1000$ and higher, substantial improvement is obtained compared to the unprotected single-qubit case.

\begin{figure}[t]
\centering
\includegraphics[width=8.6cm]{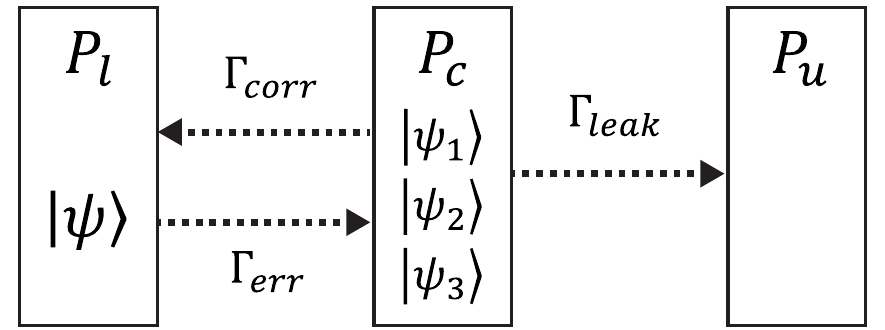}
\caption{
Rate equation model. The dynamics can be described by three subspaces and the decay rates between them. The logical subspace (population $P_l$) contains the codeword $\ket{\psi}$. It decays to the correctable subspace ($P_c$) with the single-error states $\ket{\psi_j}$ ($j=1,2,3$) at a rate $\Gamma_{\rm err}$. The error correction restores the logical subspace with a rate $\Gamma_{\rm corr}$. The single-error subspace decays to the subspace of uncorrectable states that contain more than one error (with population $P_u$) at a rate $\Gamma_{\rm leak}$.
}
\label{FigRateModel}
\end{figure}

\subsection{Rate equation model}

\label{SecRate}

We reduce the complexity of the model further by reducing the effective dynamics to a system of rate equations.
Here, we consider three subspaces:
The logical subspace, with population $P_l$, is comprised by the manifold of encoded states $\ket{\psi} = (\ket{000}+\ket{111})/\sqrt{2}$.
The correctable subspace, with population $P_c$ contains the states where one error has occurred, $\ket{\psi_j}$.
States where more than one qubit has been flipped form the subspace of uncorrectable states, $P_u$.
The described partitioning is possible due to the fact that the decay rates between the states of the considered subspaces are identical: all $\ket{\psi_j}$ decay to $\ket{\psi}$ with $\kappa_{{\rm eff},1}$, and $\ket{\psi}$ decays to $\ket{\psi_j}$ with $\kappa_{{\rm eff},3}$.
In our analysis, we assume that we operate in the regime where $G \gg \Gamma$. This means that the dominant effect which limits the driving is the reduction of the pumping strength in Eq. \eqref{EqKappaEffStrong}, and we do not need to consider the excitation of the excited states. 
We also note that our simplified model does not capture the steady state fidelities $\lim_{t\rightarrow\infty} F(t) > 0$. This is however of less concern here, where we are mainly interested in conditions for achieving a high fidelity.
With the above assumptions we obtain a system of coupled differential equations,
\begin{align*}
\dot{P}_l &= - \Gamma_{\rm err} P_l + \Gamma_{\rm corr} P_c,
\\
\dot{P}_c &= \Gamma_{\rm err} P_l - (\Gamma_{\rm corr} + \Gamma_{\rm leak}) P_c,
\\
\dot{P}_u &= \Gamma_{\rm leak} P_c,
\end{align*}
with rates $\Gamma_{\rm err} = 3 (\Gamma + \kappa_{{\rm eff},3})$, $\Gamma_{\rm corr} = \kappa_{{\rm eff},1}$ (neglecting that random bit flips at a rate $\Gamma$ can also lead back to the desired state), and $\Gamma_{\rm leak} = 2 (\Gamma + \kappa_{{\rm eff},2})$. The subspaces and transition rates are illustrated in Fig. \ref{FigRateModel}. Solving for $P_l(t)$ under the initial condition that $P_l(0)=1$ yields
\begin{align}
F(t) = P_l(t) = e^{- \Gamma_{s} t} \left[ 1 - E_0 \left( 1 - e^{- \Gamma_{f} t} \right) \right].
\label{Eq:Pl}
\end{align}
The evolution of the logical population is thus dominated by two exponential decays, a fast one with the rate
\begin{align*}
\Gamma_f = \sqrt{\Gamma_{\rm total}^2 - 4 \Gamma_{\rm err} \Gamma_{\rm leak}},
\end{align*}
where $\Gamma_{\rm total} = \Gamma_{\rm corr} + \Gamma_{\rm err} + \Gamma_{\rm leak}$ and a slow one with rate
\begin{align*}
\Gamma_s = \frac{1}{2}(\Gamma_{\rm total} - \Gamma_f) \approx \frac{\Gamma_{\rm err} \Gamma_{\rm leak}}{\Gamma_{\rm corr}}.
\end{align*}
For intermediate times $1/\Gamma_f \ll t \sim 1/\Gamma$, the exponential decays appear linear, as seen in Fig. \ref{FigCurvesApp}. The two timescales $\Gamma_f$ and $\Gamma_s$ thus generate the characteristic ``hockey stick'' form of the curve that can be seen from Fig. \ref{FigCurves}.
Assuming the drop due to $\Gamma_f$ to be faster than the timescale of interest, we can simplify Eq. \eqref{Eq:Pl} to
\begin{align}
F(t) = P_l(t) = e^{- \Gamma_{\rm eff} t} \left( 1 - E_0 \right).
\label{EqF}
\end{align}
In this simplified model, the evolution is described by the two quantities:
\begin{align*}
E_0 &= \frac{\Gamma_{\rm err} - \Gamma_{\rm eff}}{\Gamma_f} \approx \frac{\Gamma_{\rm err}}{\Gamma_{\rm corr}},
\\
\Gamma_{\rm eff} &= \Gamma_{\rm leak} E_0.
\end{align*}
Here $E_0$ denotes the static error of the scheme that arises at approximately $t = 0$, where the scheme is switched on.
The physical explanation for this initial drop is that $P_c$ and $P_l$ come to equilibrium fast and then population slowly leaks from this coupled subspace. Alternatively it can be understood from the fact that it takes a certain time to pump back after a spin flip. The effective decay rate $\Gamma_{\rm eff}$, previously $\Gamma_s$, is the slow effective decay rate of the codeword when it is protected by the error correction scheme.

\subsection{Minimization of the error and the decay rate}

\label{SecMinimization}

Using the solution of the rate equation model, we can optimize the fidelity $F = P_l(t)$ in Eq. \eqref{EqF} at a given time $t$ by the choice of available parameters $\kappa_{\rm eng}$ and $\Omega$. For that we need to fulfill the condition:
\begin{align}
\frac{\partial F}{\partial x} = - \left( \frac{\partial E_0}{\partial x} + (1-E_0) \frac{\partial \Gamma_{\rm eff}}{\partial x} t \right) e^{-\Gamma_{{\rm eff}} t} = 0,
\label{EqDF}
\end{align}
for $x \in \{\kappa_{\rm eng}, \Omega^2\}$. For the derivatives we have
\begin{align}
\frac{\kappa_{\rm eng}}{E_0} \frac{\partial E_0}{\partial \kappa_{\rm eng}}
&\approx \frac{\kappa_{\rm eng} \Omega^2}{16 \Gamma G^2 + \kappa_{\rm eng} \Omega^2} + \frac{\kappa_{\rm eng}^2 - r \Omega^2}{\kappa_{\rm eng}^2 + r \Omega^2},
\label{EqDkappa}
\\
\frac{\Omega^2}{E_0} \frac{\partial E_0}{\partial \Omega^2}
&\approx \frac{\kappa_{\rm eng} \Omega^2}{16 \Gamma G^2 + \kappa_{\rm eng} \Omega^2} - \frac{\kappa_{\rm eng}^2}{\kappa_{\rm eng}^2 + r \Omega^2},
\label{EqDOmega}
\\
\frac{x}{\Gamma_{\rm eff}} \frac{\partial \Gamma_{\rm eff}}{\partial x}
&\approx \frac{x}{E_0} \frac{\partial E_0}{\partial x} + \frac{\kappa_{\rm eng} \Omega^2}{4 \Gamma G^2 + \kappa_{\rm eng} \Omega^2}.
\label{EqDGammaEff}
\end{align}
By setting these derivatives to zero and comparing Eq. \eqref{EqDkappa} and Eq. \eqref{EqDOmega}, and Eq. \eqref{EqDGammaEff} for $x=\kappa$ and $x=\Omega^2$, we find that
\begin{align}
\Omega_{\rm opt} = \sqrt{\frac{2}{r}} \kappa_{\rm eng, opt} \approx 0.9 ~ \kappa_{\rm eng, opt}.
\label{EqOmegaOpt}
\end{align}
To solve for $\kappa_{\rm eng, opt}$, we make an ansatz $\kappa_{\rm eng, opt} \Omega_{\rm opt}^2 = \alpha \Gamma G^2$
With this, Eq. \eqref{EqDF} can be written as
\begin{align*}
(2 + 5 \tau_A) \alpha^2 + 8 \tau_A (\tau_A -1) - 64 (\tau_A + 1) = 0,
\end{align*}
where $\tau_A = A \tau_{\rm eff}$ with $A = (1-E_0)/E_0$ and $\tau_{\rm eff} = \Gamma_{\rm eff} t$. From this we derive the condition
\begin{align*}
\alpha = \frac{- 4 [ (5 \tau_A - 1) + 3 \sqrt{5 \tau_A^2 + 2 \tau_A + 1}]}{2 + 5 \tau_A}.
\end{align*}
Note that we neglected one solution because of the requirement $\alpha > 0$. The optimal decay rate is then given by
\begin{align}
\kappa_{\rm eng, opt} \approx \sqrt[3]{\frac{r ~ \alpha}{2} ~ \Gamma G^2}.
\label{EqkappaOpt}
\end{align}
Finally, this leads to the optimal static error and effective decay rate of the form
\begin{align}
E_0 &\approx \beta_E \left( \frac{\Gamma}{G} \right)^{2/3},
\label{EqE0opt}
\\
\Gamma_{\rm eff} &\approx \beta_\Gamma \left( \frac{\Gamma}{G} \right)^{2/3} \Gamma,
\label{EqGammaepsilonopt}
\end{align}
with the numeric prefactors $\beta_E$ and $\beta_\Gamma$ being given by
\begin{align*}
\beta_E(\tau_A) &= \frac{9}{16} (16 + \alpha) \left( \frac{\beta^2}{4 \alpha} \right)^{1/3},
\\
\beta_\Gamma(\tau_A) &= \frac{9}{32} (16 + \alpha)(4 + \alpha) \left( \frac{\beta^2}{4 \alpha} \right)^{1/3}.
\end{align*}
We thus find that the optimal parameters depend on the operation time of the protocol.
Here, we regard to limiting cases:
First, we consider the long-time limit where $\tau_A \gg 1$. This corresponds to optimizing $\Gamma_{\rm eff}$ under the assumption that $\Gamma_{\rm eff} t \gg E_0$. 
This yields $\alpha_\infty \approx 4(3\sqrt{5}-5)/5 \approx 1.4$ and thereby results in an optimal decay rate $\kappa_{\rm eng, opt} \approx 1.2 \sqrt{\Gamma G^2}$. For the correction rate, we obtain $\Gamma_{\rm corr} = 2 \kappa / (3 \beta) \approx 0.3 \sqrt[3]{\Gamma G^2}$, for the error and leakage rates $\Gamma_{\rm err} \approx 3.3 ~ \Gamma$ and $\Gamma_{\rm leak} \approx ~ 2.7 ~ \Gamma$.
For the prefactors in Eqs. \eqref{EqE0opt}--\eqref{EqGammaepsilonopt} we thus find $\beta_E \approx 10$ and $\beta_\Gamma \approx 22$.

Alternatively, the protocol can be optimized for short operation times, $\tau_A \rightarrow 0$. Here, minimizing the initial drop $E_0$ requires a stronger correction rate, and thus, higher $\kappa_{\rm eng}$, at the expense of a higher leakage rate. We obtain $\alpha_0 = 8$, resulting in $\kappa_{\rm eng} \approx 2.2 \sqrt[3]{\Gamma G^2}$, $\Gamma_{\rm corr} \approx 0.6 \sqrt[3]{\Gamma G^2}$, $\Gamma_{\rm err} = 9/2 ~ \Gamma$, and $\Gamma_{\rm leak} = 6 ~ \Gamma$. This yields the coefficients $\beta_E \approx 8$ and $\beta_\Gamma = 47$. The reduction of the initial drop by this parameter choice is therefore not very pronounced ($\beta_E=8$ vs. $\beta_E=10$), whereas the slope of the effective decay is strongly increased ($\beta_\Gamma=22$ vs. $\beta_\Gamma=47$). The optimization for $\tau_A \gg 1$ hence constitutes the better choice and is used for the metrology application.

The above findings for $\kappa_{\rm eng, opt}$ in the long-time limit are in good agreement with the numerical optimization of this parameter from the truncated dynamics which fulfills $\tau_A \gg 1$.
Using $\beta = 2.5$, the optimal driving strength in Eq. \eqref{EqOmegaOpt} is, however, slightly stronger than the result obtained from the numerical optimization, where we find $\Omega_{\rm opt} \approx 4 \kappa_{\rm eng, opt}/5$. This is due to the fact that we ignored the population in the excited states: This amounts to about $4 - 7 \%$ at $t = 1 / \Gamma$ for typical parameters and would thus yield a smaller value for $\Omega_{\rm opt}$. The resulting discrepancy vanishes, however, in the limit of large $G / \Gamma$.

\begin{figure}[t]
\centering
\includegraphics[width=8.6cm]{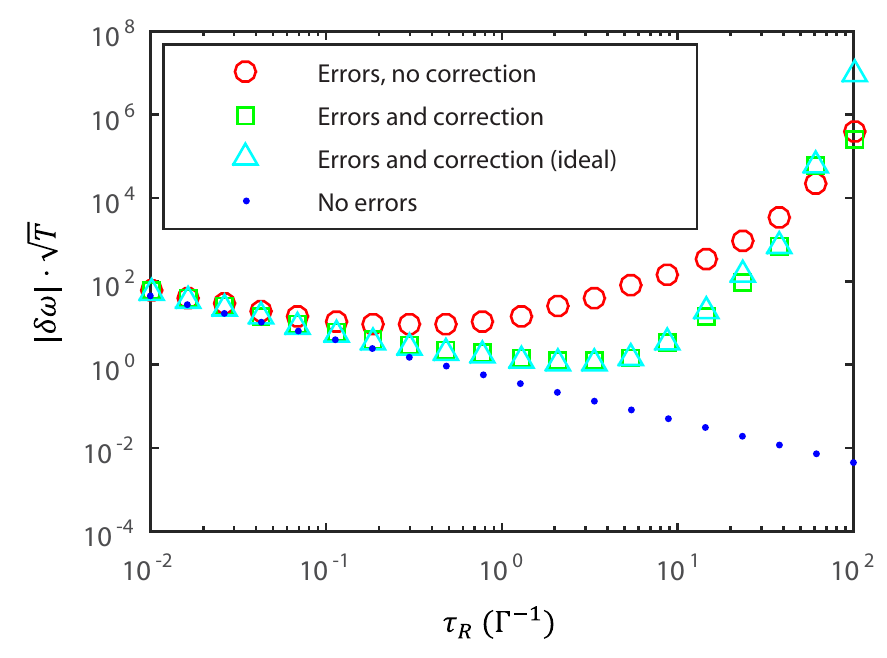}
\caption{
Quantum error correction enhanced sensitivity in the presence of parallel noise. The plots show the normalized measurement precision $|\delta\omega|\sqrt{T}$ of a spin-flip corrected Ramsey measurement in the presence of additional phase noise.
As a consequence of the uncorrected noise, the prolongation of the Ramsey time is less effective for longer times $\Gamma \tau_{\rm R} \gtrsim 10^{2}$, but still pronounced for $\Gamma \tau_{\rm R} \sim 10^{0}$. In the simulation, we use $G = 5000 ~ \Gamma$, $\Gamma_Z = \Gamma / 100$, where $\Gamma_Z$ ($\Gamma$) is the collective phase flip (individual bit flip) rate, and the optimized parameters $\Omega = 4 \kappa_{\rm eng}/5$ and $\kappa_{\rm eng} = 1.2 \sqrt[3]{\Gamma G^2}$.
}
\label{FigSensitivityApp}
\end{figure}

\section{Application to quantum metrology}\label{App_Metrology}

In the following, we explain the sensitivity calculation underlying the numerical results in Figs.~\ref{FigMetrology}, \ref{FigSensitivity}, and \ref{FigSensitivityApp} (App.~\ref{App_Metrology_Sensitivity}) and describe how our quantum error correction enhanced sensing scheme can be scaled up to larger particle numbers (App.~\ref{App_Metrology_Scalability}).

\subsection{Sensitivity of the measurement}\label{App_Metrology_Sensitivity}

First we review how the sensitivity of a Ramsey spectroscopy scheme with entangled particles can be determined~\cite{ShotNoise, MetrologyWineland1, MetrologyWineland2, Huelga}. 
Using the Ramsey sequence described in the main text, a first Ramsey pulse prepares a system of $N$ qubits in the state 
$\ket{\psi(0)}^{\otimes N}=(\ket{0}^{\otimes N}+\ket{1}^{\otimes N})/\sqrt{2}$, which evolves under the Hamiltonian $H=(\omega/2)\sum_{j=1}^N\sigma_j^z$ for a time $\tau_{\rm R}$, resulting in the state
$\ket{\psi(\tau_{\rm R})}=(\ket{0}^{\otimes N}+e^{-i N \omega\tau_{\rm R}}\ket{1}^{\otimes N})/\sqrt{2}$.
After the second Ramsey pulse one of the qubits is measured. The probability to find this qubit in state $\ket{1}$ is given by $P_1=\cos^2(N\omega\tau_{\rm R}/2)$.
The uncertainty in estimating $P_1$ due to the statistical fluctuations associated with a finite sample 
$\Delta P_1= \sqrt{P_1(1-P_1)/n_{\rm data}}$, 
depends on the number of experimental data $n_{\rm data}$. In the considered case this is given by the number of runs $n_{\rm runs}=n_{\rm data}=T/\tau_{\rm R}$, where $T$ is the total measurement time. The uncertainty in the measurement of $\omega$ is therefore given by
$$|\delta\omega|=\frac{\Delta P_1}{|dP_1/d\omega |}.$$
In the ideal case, this yields $|\delta\omega|=1/(N\sqrt{T\tau_R})$. The numerical results in Figs.~\ref{FigMetrology}, \ref{FigSensitivity}, and \ref{FigSensitivityApp} have been obtained by calculating the time evolution of the density matrix of a three-ion system $\rho(t)$ in the presence of the signal Hamiltonian, individual bit flips and the error-correcting scheme. Fig. \ref{FigSensitivityApp} shows the dynamics that is obtained if collective dephasing is added to the problem.

\subsection{Extension to $N_{\rm log}$ logical qubits}\label{App_Metrology_Scalability}

Here we consider an extension of the metrology scheme discussed in the main text to $N_{\rm log}$ logical qubits, using a segmented trap~\cite{SegmentedTrap1,SegmentedTrap2}. In this setup, groups of three ions constitute logical qubits that are trapped in separate trap segments. For sufficiently high potential barriers separating the ion triples, the motional modes associated with the different logical qubits can be assumed to be independent. The Ramsey sequence explained in the main text can be straightforwardly generalized to this setting, as explained in the following.\\
\\
\underline{Ramsey scheme using a product state of $N_{\rm log}$ logical qubits:}
\begin{enumerate}[label=(\roman*)]
\item Starting from the initial state $\ket{0}^{\otimes N_{\rm log}}_L=\ket{000}^{\otimes N_{\rm log}}$, individual Ramsey pulses are applied to the logical qubits to prepare the state $\ket{\psi(0)} =|+\rangle_L^{\otimes N_{\rm log}}$ with $|+\rangle_L=(\ket{000}+ \ket{111})/\sqrt{2}$. 
\item  During the Ramsey waiting time of duration $\tau_{\rm{R}}$, each logical qubit state picks up a relative phase $\phi(\tau_{\rm{R}})=3\omega \tau_{\rm{R}}$, such that $\ket{\psi(\tau_{\rm{R}})} =\left( (\ket{000} + e^{-i 3\omega\tau_{\rm{R}}} \ket{111})/\sqrt{2}\right)^{\otimes N_{\rm log}}$.
\item A second set of individual Ramsey pulses that acts on each logical qubit separately, transforms the evolved state into $\ket{\psi'(\tau_{\rm{R}})} =\left(\cos(3\omega\tau_{\rm{R}}/2)\ket{111}+i\sin(3\omega\tau_{\rm{R}}/2)\ket{000}\right)^{\otimes N_{\rm log}}$. 
\item For each logical qubit (i.e. for each trap segment), a measurement on one of the three physical qubits is performed in the $\sigma^z$ basis. For each ion triple, the probability to detect the first (or any other) qubit in state $\ket{1}$ is given by $P_1=\cos^2(3\omega\tau_{\rm{R}}/2)$, which allows one to determine the phase $\phi(\tau_{\rm{R}})$ and therefore the signal strength $\omega$.
\end{enumerate}
In the ideal case, the sensitivity of this measurement \cite{Ndata} is given by $|\delta \omega|=1/(3\sqrt{N_{\rm log} \tau_{\rm{R}} T})$, where $T=n_{\rm runs}\tau_{\rm R}$ is the total measurement time. As described in the main text, the dependence of the measurement precision $|\delta \omega|$ on the Ramsey time $\tau_{\rm R}$ changes in the presence of noise. The limitations due transversal noise can be mitigated by applying our quantum error correction scheme. Since the protocol is applied for each trap segment individually, the results discussed in the main text generalize directly to the setting involving $N_{\rm log}$ logical qubits.\\
\\The Ramsey sequence described above does not yield a quantum enhancement of the scaling of the measurement precision $|\delta \omega|$ with in the number of logical qubits $N_{\rm log}$ since the logical qubits remain in a product state throughout the measurement sequence. To conclude this section, we consider an alternative setting where collective effects with respect to the logical qubits play a role and where quantum error correction allows one in principle to perform Heisenberg limited precision measurements even in the presence of noise. This setting is conceptually interesting even though its applicability may be limited in practice. More specifically, we consider the following Ramsey sequence assuming a trap with adjustable segmentation.\\ 
\\\underline{Ramsey scheme using $N_{\rm log}$ entangled logical qubits:}
\begin{enumerate}[label=(\roman*)]
\item The ions are initially held in an unsegmented trap and couple to a common motional mode which allows one to apply a global Ramsey pulse transforming the initial state $\ket{0}^{\otimes 3N_{\rm{log}}}_L$ into $\ket{\psi(0)} =(\ket{0}^{\otimes 3N_{\rm{log}}}+\ket{1}^{\otimes 3N_{\rm{log}}})/\sqrt{2}$. 
\item  After the first Ramsey pulse, potential barriers are ramped up which divide the trap into $N_{\rm{log}}$ independent trap segments that each contain a logical qubit. This segmentation into sets of three ions is kept during the whole Ramsey waiting period $\tau_{\rm{R}}$ and the quantum error correction scheme is applied for each segment individually. Ideally, the relative phase acquired during the Ramsey waiting time results in $\ket{\psi(\tau_{\rm{R}})}=(\ket{0}^{\otimes 3N}+e^{-i3N_{\rm{log}}\omega\tau_{\rm{R}}} \ket{1}^{\otimes 3N})/\sqrt{2}$.
\item After the Ramsey waiting period, the potential barriers of the trap are ramped down such that the ions share again a motional mode that  can be used to perform a second global Ramsey pulse resulting in the state $\ket{\psi'(\tau_{\rm{R}})} =\cos(3N_{\rm{log}}\omega\tau_{\rm{R}}/2)\ket{1}^{\otimes 3N_{\rm{log}}}+i\sin(3N_{\rm{\log}}\omega\tau_{\rm{R}}/2)\ket{0}^{\otimes 3N_{\rm{log}}}$. 
\item As last step, one of the $3N_{\rm{\log}}$ qubits is measured in the $\sigma^z$ basis.
\end{enumerate}
In the absence of imperfections, the sensitivity is given by $\sqrt{T}|\delta \omega|=1/(N\sqrt{\tau_{\rm{R}}})$, where the $N=3 N_{\rm log}$ is the number of ions.
If we consider a situation involving only the signal Hamiltonian and spin flips at a rate $\Gamma$ (in the absence of error correction), the measurement precision drops to $\sqrt{T}|\delta \omega|= c \Gamma^{1/6} / N^{5/6}$, where $c$ is a constant numerical factor (see~\cite{Acin}).
If a continuous three-qubit repetition code is operated in the strong correction regime where the correction rate $\Gamma_{\rm corr}$ is much larger than the bit flip rate $\Gamma$, the dynamics can be described in terms of logical qubits that undergo logical bit flips $\ket{000}\leftrightarrow\ket{111}$ at a reduced rate $\Gamma_{\rm{L}}$. In this regime, the population of states involving one or two single-qubit bit flips becomes negligible~\cite{Ippoliti2014}. As a result, the achievable measurement precision is given by $\sqrt{T}|\delta \omega|\propto \Gamma_{\rm{L}}^{1/6} / N^{5/6}$, where $\Gamma_L\propto\epsilon\Gamma$ with $\epsilon=\Gamma/\Gamma_{\rm corr}$. By scaling up the correction rate with the number of qubits such that $\epsilon \propto N^{-1/6}$, the measurement precision of the scheme is Heisenberg limited (up to a maximum number of logical qubits $N_{\rm log, max}$ that is determined by the maximum achievable error correction rate). While it is conceptually interesting that the Heisenberg scaling can be restored in principle, experimental imperfections will considerably limit its applicability. Still, the Ramsey scheme based on a product state of $N_{\rm log}$ logical qubits described above is likely to be of higher practical value for current quantum hardware.


\begin{thebibliography}{99}
\bibitem{LambDicke}
The Lamb-Dicke approximation holds for $\eta^2(2\bar{n}+1)\ll1$, where $\bar{n}$ is the average phonon number and $\eta\!=\!2 \pi x_0/\lambda$ is the Lamb-Dicke parameter given by the ratio of the ion's ground state wave packet size $x_0$ and the wavelength of the applied light field $\lambda$.  For the considered type of experiment, we consider the Lamb-Dicke parameter to be small.
\bibitem{Ndata}
In the case considered here, the number of experimental data is given by $n_{\rm data} = N_{\rm{log}} \cdot n_{\rm runs} = N_{\rm{log}} T/\tau_{R}$. Evaluating $\Delta P_1$ and $dP_1/d\omega$ as described in App.~\ref{App_Metrology_Sensitivity} yields therefore $|\delta \omega|=1/(3\sqrt{N_{\rm log} \tau_{\rm{R}} T})$.
\end{thebibliography}
\end{document}